\documentclass[11pt]{article}
\usepackage{amssymb,amsfonts,amsmath,graphicx,hyperref}

\newcommand{\mcm}[3]{\newcommand{#1}[#2]{{\ensuremath{#3}}}}

\usepackage{latexsym}
\usepackage{amssymb}

\mcm{\blank}{0}{(\emptybk)} \mcm{\dashbk}{0}{\mbox{---}}
\mcm{\emptybk}{0}{\:\:} \mcm{\hyph}{0}{\mbox{-}}

\mcm{\diagspace}{0}{\mbox{\hspace{2em}}}

\mcm{\cat}{1}{\mc{#1}} \mcm{\fcat}{1}{\mb{#1}}
\mcm{\mc}{1}{\mathcal{#1}} \mcm{\mr}{1}{\mathrm{#1}}
\mcm{\mi}{1}{\mathit{#1}} \mcm{\mb}{1}{\mathbf{#1}}
\mcm{\scat}{1}{\Bbb{#1}} \mcm{\twid}{1}{\widetilde{#1}}

\mcm{\elt}{0}{\in} \mcm{\sub}{0}{\,\subseteq\,}
\mcm{\such}{0}{\:|\:} \mcm{\without}{0}{\setminus}

\mcm{\atsr}{0}{\Box} \mcm{\eqv}{0}{\,\simeq\,}
\mcm{\iso}{0}{\,\cong\,}
\mcm{\of}{0}{\raisebox{0.2mm}{\ensuremath{\scriptstyle\circ}}}

\mcm{\bdry}{0}{\partial}

\mcm{\Bee}{0}{\cat{B}} \mcm{\Beep}{0}{\cat{B'}}
\mcm{\Eee}{0}{\cat{E}} \mcm{\Eeep}{0}{\cat{E'}}

\mcm{\Ess}{0}{\cat{S}} \mcm{\Tee}{0}{\cat{T}}
\mcm{\Teep}{0}{\cat{T'}} \mcm{\Stee}{0}{\scat{T}}
\mcm{\Steep}{0}{\scat{T'}}

\mcm{\blbk}{0}{\blank^{\blob}}
\mcm{\blob}{0}{\scriptscriptstyle{\bullet}}
\mcm{\stbk}{0}{\blank^{*}} \mcm{\ubl}{0}{{}^{\blob}}
\mcm{\ust}{0}{{}^{*}}

\mcm{\Cartpr}{0}{\pr{\Eee}{T}} \mcm{\Cartprp}{0}{\pr{\Eeep}{T'}}
\mcm{\Mnd}{0}{\triple{T}{\eta}{\mu}}
\mcm{\Zeropr}{0}{\pr{\Set}{\id}}

\mcm{\dopset}{0}{\ftrcat{\Delta^{\op}}{\Set}}
\mcm{\tropset}{0}{\ftrcat{\fcat{TR}^{\op}}{\Set}}

\mcm{\cod}{0}{\mr{cod}} \mcm{\dom}{0}{\mr{dom}}
\mcm{\End}{0}{\mr{End}} \mcm{\Hom}{0}{\mr{Hom}}
\mcm{\ob}{0}{\mr{ob}\,} \mcm{\op}{0}{\mr{op}}

\mcm{\comp}{0}{\mi{comp}} \mcm{\id}{0}{\mi{id}}
\mcm{\ids}{0}{\mi{ids}} \mcm{\mult}{0}{\mi{mult}}
\mcm{\unit}{0}{\mi{unit}}

\mcm{\Ab}{0}{\fcat{Ab}} \mcm{\Alg}{0}{\fcat{Alg}}
\mcm{\Bim}{1}{\fcat{Bim}(#1)} \mcm{\Cat}{0}{\fcat{Cat}}
\mcm{\Cay}{0}{\fcat{Cay}} \mcm{\Cpn}{1}{\pr{\Set/S_{#1}}{T_{#1}}}
\mcm{\fc}{0}{\fcat{fc}} \mcm{\fm}{0}{\fcat{fm}}
\mcm{\Graph}{0}{\fcat{Graph}} \mcm{\Gy}{0}{\fcat{Gy}}
\mcm{\Hpn}{1}{\pr{\Eee_{#1}}{P_{#1}}} \mcm{\Mon}{0}{\mb{Mon}}
\mcm{\Multicat}{0}{\fcat{Multicat}} \mcm{\One}{0}{\fcat{1}}
\mcm{\PD}{1}{\fcat{PD}_{#1}} \mcm{\Prof}{0}{\fcat{Prof}}
\mcm{\Set}{0}{\fcat{Set}} \mcm{\Span}{0}{\fcat{Span}}
\mcm{\Ssq}{0}{\fcat{Ssq}} \mcm{\Struc}{0}{\fcat{Struc}}
\mcm{\Sym}{0}{\fcat{Sym}} \mcm{\TR}{1}{\fcat{TR}(#1)}
\mcm{\Tr}{0}{\fcat{Tr}} \mcm{\Twocat}{0}{\fcat{2\hyph\Cat}}

\mcm{\integers}{0}{\mathbb{Z}}

\mcm{\range}{2}{#1,\,\ldots\,,#2}
\mcm{\bftuple}{2}{\tuplebts{\range{#1}{#2}}}
\mcm{\tuple}{3}{\tuplebts{\range{#1,#2}{#3}}}
\mcm{\rttuple}{1}{\tuplebts{\,\ldots\,,#1}}
\mcm{\abftuple}{2}{\atuplebts{\range{#1}{#2}}}
\mcm{\atuple}{3}{\atuplebts{\range{#1,#2}{#3}}}
\mcm{\arttuple}{1}{\atuplebts{\,\ldots\,,#1}}
\mcm{\sqbftuple}{2}{\obt\range{#1}{#2}\cbt}
\mcm{\pr}{2}{\tuplebts{#1,#2}}
\mcm{\triple}{3}{\tuplebts{#1,#2,#3}}

\mcm{\eend}{2}{#1[#2]} \mcm{\ehom}{3}{#1[#2,#3]}
\mcm{\ftrcat}{2}{[#1,#2]} \mcm{\homset}{3}{#1(#2,#3)}
\mcm{\multihom}{3}{#1(#2;#3)}
\mcm{\relhom}{5}{#1_{#2}(\range{#3}{#4};#5)}

\mcm{\go}{0}{\rTo} \mcm{\goby}{1}{\rTo^{#1}}
\mcm{\goesto}{0}{\,\longmapsto\,} \mcm{\goiso}{0}{\goby{\diso}}
\mcm{\monic}{0}{\rMonic} \mcm{\og}{0}{\lTo}
\mcm{\ogby}{1}{\lTo^{#1}}

\mcm{\gph}{2}{\spn{#1}{T #2}{#2}} \mcm{\graph}{4}{\spaan{#1}{T
#2}{#2}{#3}{#4}} \mcm{\oppair}{2}{\stackrel{\rTo^{#1}}{\lTo_{#2}}}
\mcm{\parpair}{2}{\stackrel{\rTo^{#1}}{\rTo_{#2}}}
\mcm{\spn}{3}{#2 \og #1 \go #3} \mcm{\spaan}{5}{#2 \ogby{#4} #1
\goby{#5} #3}

\mcm{\bktdvslob}{3}
    {\left(
    \begin{diagram}[height=1.5em]
    #1      \\
    \dTo>{\,#2} \\
    #3      \\
    \end{diagram}
    \right)}
\mcm{\slob}{3}{(#1 \goby{#2} #3)} \mcm{\vslob}{3}
    {\left.
    \begin{diagram}[height=1.5em]
    #1      \\
    \dTo>{\,#2} \\
    #3      \\
    \end{diagram}
    \right.}

\newenvironment{tree}
    {\begin{diagram}[height=1em,width=.75em,abut,noPS,tight]}
    {\end{diagram}}

\mcm{\enode}{0}{\circ}

\mcm{\nl}{1}{\stackrel{\textstyle #1}{\node}}
\mcm{\node}{0}{\bullet}

\mcm{\utree}{0}{\node}

\mcm{\diso}{0}{\sim}

\mcm{\vdiso}{0}{\wr}

\mcm{\nat}{0}{\mathbb{N}}

\mcm{\Onepr}{0}{\pr{\Graph}{\fc}}
\newlength{\nllwidth}
\newlength{\nllheight}
\newcommand{\stackbelow}[2]{%
\settowidth{\nllwidth}{\ensuremath{#1}\ensuremath{#2}}%
\settoheight{\nllheight}{\ensuremath{#2}}%
\addtolength{\nllheight}{.3ex}%
\mbox{%
\ensuremath{#1}%
\hspace{-.5\nllwidth}%
\raisebox{-1\nllheight}{\ensuremath{#2}}}}
\mcm{\nlal}{2}{\stackbelow{\nl{#1}}{#2}}
\mcm{\nll}{1}{\stackbelow{\node}{#1}} \mcm{\wun}{0}{\fcat{1}}
\mcm{\atuplebts}{1}{\langle #1 \rangle} \mcm{\tuplebts}{1}{(#1)}
\mcm{\bo}{0}{(} \mcm{\bc}{0}{)}
\mcm{\UBilax}{0}{\fcat{UBicat}_\mr{lax}}
\mcm{\UBiwk}{0}{\fcat{UBicat}_\mr{wk}}
\mcm{\UBistr}{0}{\fcat{UBicat}_\mr{str}}
\mcm{\Bilax}{0}{\fcat{Bicat}_\mr{lax}}
\mcm{\Biwk}{0}{\fcat{Bicat}_\mr{wk}}
\mcm{\Bistr}{0}{\fcat{Bicat}_\mr{str}} \mcm{\rotsub}{0}{\cup
\raisebox{0.1em}{$\scriptstyle{|}$}} \mcm{\pd}{0}{\fcat{pd}}
\mcm{\rep}{1}{\widehat{#1}} \mcm{\ovln}{1}{\overline{#1}}
\mcm{\Gph}{0}{\fcat{Gph}} \mcm{\tr}{0}{\fcat{tr}}

\mcm{\ladj}{0}{\,\dashv\,} \mcm{\zeropd}{0}{\node}
    {\end{diagram}}
\mcm{\END}{0}{\fcat{End}} \mcm{\HOM}{0}{\fcat{Hom}}

\newlength{\gwidth} 
\newlength{\gvert}  
\newlength{\gdrop}  
\newlength{\gbaredrop}  
\newlength{\goffset}    
\newlength{\gtemp}  

\newcommand{\present}[1]{%
\makebox[1\gwidth]{%
\rule[-1\gdrop]{0ex}{1\gvert}%
\raisebox{-1\gbaredrop}{#1}}}

\newcommand{\presentl}[1]{%
\makebox[1\gwidth][l]{%
\rule[-1\gdrop]{0ex}{1\gvert}%
\raisebox{-1\gbaredrop}{#1}}}

\newcommand{\presentr}[1]{%
\makebox[1\gwidth][r]{%
\rule[-1\gdrop]{0ex}{1\gvert}%
\raisebox{-1\gbaredrop}{#1}}}

\newcommand{\ginitdims}[2]{
\setlength{\unitlength}{1em}
\setlength{\goffset}{.25\unitlength}
\setlength{\gwidth}{#1\unitlength}
\setlength{\gvert}{#2\unitlength}
\setlength{\gdrop}{.5\gvert}
\addtolength{\gdrop}{-1\goffset}
\setlength{\gbaredrop}{1\gdrop}
\addtolength{\gvert}{.6\unitlength}
\addtolength{\gdrop}{.3\unitlength}}    

\newcommand{\cinitdims}[2]{
\setlength{\unitlength}{1em}
\setlength{\goffset}{.35\unitlength}
\setlength{\gwidth}{#1\unitlength}
\setlength{\gvert}{#2\unitlength}
\setlength{\gdrop}{.5\gvert}
\addtolength{\gdrop}{-1\goffset}
\setlength{\gbaredrop}{1\gdrop}
\addtolength{\gvert}{.6\unitlength}
\addtolength{\gdrop}{.3\unitlength}}    

\newcommand{\gsinitdims}[2]{
\setlength{\unitlength}{0.5em}
\setlength{\goffset}{.25\unitlength}
\setlength{\gwidth}{#1\unitlength}
\setlength{\gvert}{#2\unitlength}
\setlength{\gdrop}{.5\gvert}
\addtolength{\gdrop}{-1\goffset}
\setlength{\gbaredrop}{1\gdrop}
\addtolength{\gvert}{.6\unitlength}
\addtolength{\gdrop}{.3\unitlength}}    

\newcommand{\sidespic}[1]{%
\settowidth{\gtemp}{\ensuremath{#1}}%
\addtolength{\gwidth}{1\gtemp}}

\newcommand{\abovepic}[1]{%
\settoheight{\gtemp}{\ensuremath{#1}}%
\addtolength{\gvert}{1\gtemp}%
\settodepth{\gtemp}{\ensuremath{#1}}%
\addtolength{\gvert}{1\gtemp}}

\newcommand{\belowpic}[1]{%
\settoheight{\gtemp}{\ensuremath{#1}}%
\addtolength{\gvert}{1\gtemp}%
\addtolength{\gdrop}{1\gtemp}%
\settodepth{\gtemp}{\ensuremath{#1}}%
\addtolength{\gvert}{1\gtemp}%
\addtolength{\gdrop}{1\gtemp}}

\newcommand{\cell}[4]{\put(#1,#2){\makebox(0,0)[#3]{\ensuremath{#4}}}}
\mcm{\zmark}{0}{\scriptstyle{\bullet}}

\newcommand{\pregfst}[1]{%
\begin{picture}(0.5,0.2)(-0.5,-0.2)%
\cell{-0.1}{-0.2}{tr}{#1}%
\cell{0}{0}{c}{\zmark}%
\end{picture}}

\mcm{\gfst}{1}{%
\ginitdims{0.5}{0.4}%
\sidespic{#1}%
\belowpic{#1}%
\presentr{\pregfst{#1}}}

\newcommand{\preglst}[1]{%
\begin{picture}(0.5,0.2)(0,-0.2)%
\cell{0.1}{-0.2}{tl}{#1}%
\cell{0.05}{0}{c}{\zmark}%
\end{picture}}

\mcm{\glst}{1}{%
\ginitdims{.5}{.4}%
\sidespic{#1}%
\belowpic{#1}%
\presentl{\preglst{#1}}}

\newcommand{\preglft}[1]{%
\begin{picture}(0,0.2)(0,-0.2)%
\cell{-0.1}{-0.2}{tr}{#1}%
\cell{0.05}{0}{c}{\zmark}%
\end{picture}}

\mcm{\glft}{1}{%
\ginitdims{0}{.4}%
\belowpic{#1}%
\present{\preglft{#1}}}

\newcommand{\pregrgt}[1]{%
\begin{picture}(0,0.2)(0,-0.2)%
\cell{0.1}{-0.2}{tl}{#1}%
\cell{0.05}{0}{c}{\zmark}%
\end{picture}}

\mcm{\grgt}{1}{%
\ginitdims{0}{.4}%
\belowpic{#1}%
\present{\pregrgt{#1}}}

\newcommand{\pregblw}[1]{%
\begin{picture}(0,0.3)(0,-0.3)
\cell{0}{-0.3}{t}{#1}%
\cell{0.05}{0}{c}{\zmark}%
\end{picture}}

\mcm{\gblw}{1}{%
\ginitdims{0}{.6}%
\belowpic{#1}%
\present{\pregblw{#1}}}

\newcommand{\pregfbw}[1]{%
\begin{picture}(0,0.65)(0,-0.65)
\cell{0}{-0.65}{t}{#1}%
\cell{0.05}{0}{c}{\zmark}%
\end{picture}}

\mcm{\gfbw}{1}{%
\ginitdims{0}{1.3}%
\belowpic{#1}%
\present{\pregfbw{#1}}}

\newcommand{\pregzero}[1]{%
\begin{picture}(0.8,0.4)(-0.4,-0.4)
\cell{0}{-0.4}{t}{#1}%
\cell{0}{0}{c}{\zmark}%
\end{picture}}

\mcm{\gzero}{1}{%
\ginitdims{0.8}{.6}%
\belowpic{#1}%
\sidespic{#1}%
\present{\pregzero{#1}}}

\newcommand{\pregone}[1]{%
\begin{picture}(5,0.4)(0,-0.2)%
\cell{2.5}{0.2}{b}{#1}%
\put(0,0){\vector(1,0){5}}%
\end{picture}}

\mcm{\gone}{1}{%
\ginitdims{5}{0.4}%
\abovepic{#1}%
\present{\pregone{#1}}}

\newcommand{\pregtwo}[3]{%
\begin{picture}(5,3.4)(0,-0.2)%
\cell{2.5}{3.2}{b}{#1}%
\cell{2.5}{-.2}{t}{#2}%
\cell{2.7}{1.5}{l}{#3}%
\qbezier(0,1.5)(2.5,4.5)(5,1.5)%
\qbezier(0,1.5)(2.5,-1.5)(5,1.5)%
\put(5,1.5){\vector(1,-1){0}}%
\put(5,1.5){\vector(1,1){0}}%
\put(2.5,2.5){\vector(0,-1){2}}%
\end{picture}}

\mcm{\gtwo}{3}{%
\ginitdims{5}{3.4}%
\abovepic{#1}%
\belowpic{#2}%
\present{\pregtwo{#1}{#2}{#3}}}

\newcommand{\pregthree}[5]{%
\begin{picture}(5,5.4)(0,-1.2)%
\cell{2.5}{4.2}{b}{#1}%
\cell{1.5}{1.7}{b}{#2}%
\cell{2.5}{-1.2}{t}{#3}%
\cell{2.7}{2.75}{l}{#4}%
\cell{2.7}{0.25}{l}{#5}%
\qbezier(0,1.5)(2.5,6.5)(5,1.5)%
\qbezier(0,1.5)(2.5,-3.5)(5,1.5)%
\put(0,1.5){\vector(1,0){5}}%
\put(2.5,3.5){\vector(0,-1){1.5}}%
\put(2.5,1){\vector(0,-1){1.5}}%
\put(5,1.5){\vector(1,-3){0}}%
\put(5,1.5){\vector(1,3){0}}%
\end{picture}}

\mcm{\gthree}{5}{%
\ginitdims{5}{5.4}%
\abovepic{#1}%
\belowpic{#3}%
\present{\pregthree{#1}{#2}{#3}{#4}{#5}}}

\newcommand{\pregfour}[7]{%
\begin{picture}(5,8.4)(0,-2.7)%
\cell{2.5}{5.7}{b}{#1}%
\cell{1.5}{2.8}{b}{#2}%
\cell{1.5}{0.2}{t}{#3}%
\cell{2.5}{-2.7}{t}{#4}%
\cell{2.7}{4.25}{l}{#5}%
\cell{2.7}{1.5}{l}{#6}%
\cell{2.7}{-1.25}{l}{#7}%
\qbezier(0,1.5)(2.5,9.5)(5,1.5)%
\qbezier(0,1.5)(2.5,4)(5,1.5)%
\qbezier(0,1.5)(2.5,-1)(5,1.5)%
\qbezier(0,1.5)(2.5,-6.5)(5,1.5)%
\put(2.5,5.25){\vector(0,-1){2}}%
\put(2.5,2.5){\vector(0,-1){2}}%
\put(2.5,-0.25){\vector(0,-1){2}}%
\put(5,1.5){\vector(1,-4){0}}%
\put(5,1.5){\vector(4,-3){0}}%
\put(5,1.5){\vector(4,3){0}}%
\put(5,1.5){\vector(1,4){0}}%
\end{picture}}

\mcm{\gfour}{7}{%
\ginitdims{5}{8.4}%
\abovepic{#1}%
\belowpic{#4}%
\present{\pregfour{#1}{#2}{#3}{#4}{#5}{#6}{#7}}}

\newcommand{\pregthreecell}[5]{%
\begin{picture}(8,5)(-4,-2.5)%
\cell{0}{2.5}{b}{#1}%
\cell{0}{-2.5}{t}{#2}%
\cell{-1.7}{0}{r}{#3}%
\cell{1.7}{0}{l}{#4}%
\cell{0}{0.2}{b}{#5}%
\qbezier(-4,0)(0,4.2)(4,0)%
\qbezier(-4,0)(0,-4.2)(4,0)%
\qbezier(-0.5,1.8)(-2.5,0)(-0.5,-1.8)%
\qbezier(0.5,1.8)(2.5,0)(0.5,-1.8)%
\put(-1,0){\vector(1,0){2}}%
\put(4,0){\vector(1,-1){0}}%
\put(4,0){\vector(1,1){0}}%
\put(-0.5,-1.8){\vector(1,-1){0}}%
\put(0.5,-1.8){\vector(-1,-1){0}}%
\end{picture}}

\mcm{\gthreecell}{5}{%
\ginitdims{8}{5}%
\abovepic{#1}%
\belowpic{#2}%
\present{\pregthreecell{#1}{#2}{#3}{#4}{#5}}}

\newcommand{\pregthreecellu}{%
\begin{picture}(5,3.4)(-0.5,-0.2)%
\qbezier(-.5,1.5)(2,4.5)(4.5,1.5)%
\qbezier(-.5,1.5)(2,-1.5)(4.5,1.5)%
\qbezier(1.5,2.7)(0.5,1.5)(1.5,0.3)%
\qbezier(2.5,2.7)(3.5,1.5)(2.5,0.3)%
\put(1.3,1.5){\vector(1,0){1.4}}%
\put(4.5,1.5){\vector(1,-1){0}}%
\put(4.5,1.5){\vector(1,1){0}}%
\put(1.5,0.3){\vector(2,-3){0}}%
\put(2.5,0.3){\vector(-2,-3){0}}%
\end{picture}}

\mcm{\gthreecellu}{0}{%
\ginitdims{5}{3.4}%
\present{\pregthreecellu}}

\newcommand{\pregtwocentre}[3]{%
\begin{picture}(5,3.4)(0,-0.2)%
\cell{2.5}{3.2}{b}{#1}%
\cell{2.5}{-.2}{t}{#2}%
\cell{2.5}{1.5}{c}{#3}%
\qbezier(0,1.5)(2.5,4.5)(5,1.5)%
\qbezier(0,1.5)(2.5,-1.5)(5,1.5)%
\put(5,1.5){\vector(1,-1){0}}%
\put(5,1.5){\vector(1,1){0}}%
\put(2.5,2.5){\vector(0,-1){2}}%
\end{picture}}

\mcm{\gtwocentre}{3}{%
\ginitdims{5}{3.4}%
\abovepic{#1}%
\belowpic{#2}%
\present{\pregtwocentre{#1}{#2}{#3}}}

\newcommand{\pregspecialone}[9]{%
\begin{picture}(8,8)(-4,-4)%
\cell{0}{3.9}{b}{#1}%
\cell{-2}{-0.2}{t}{#2}%
\cell{0}{-3.9}{t}{#3}%
\cell{-1.5}{1.1}{r}{#4}%
\cell{0.2}{1.5}{l}{#5}%
\cell{1.5}{1.1}{l}{#6}%
\cell{0.2}{-2}{l}{#7}%
\cell{-0.9}{2.3}{b}{#8}%
\cell{0.9}{2.3}{b}{#9}%
\qbezier(-4,0)(0,8)(4,0)%
\qbezier(-4,0)(0,-8)(4,0)%
\qbezier(-0.5,3.4)(-3.5,2)(-0.5,0.6)%
\qbezier(0.5,3.4)(3.5,2)(0.5,0.6)%
\put(-4,0){\vector(1,0){8}}%
\put(0,3.4){\vector(0,-1){2.8}}%
\put(0,-0.8){\vector(0,-1){2.4}}%
\put(-1.5,2.2){\vector(1,0){1.2}}%
\put(0.3,2.2){\vector(1,0){1.2}}%
\put(4,0){\vector(1,-2){0}}%
\put(4,0){\vector(1,2){0}}%
\put(-0.5,0.6){\vector(2,-1){0}}%
\put(0.5,0.6){\vector(-2,-1){0}}%
\end{picture}}

\mcm{\gspecialone}{9}{%
\ginitdims{8}{8}%
\abovepic{#1}%
\belowpic{#3}%
\present{\pregspecialone{#1}{#2}{#3}{#4}{#5}{#6}{#7}{#8}{#9}}}

\newcommand{\pregspecialtwo}{%
\begin{picture}(5,3.4)(0,-0.2)%
\qbezier(0,1.5)(2.5,4.5)(5,1.5)%
\qbezier(0,1.5)(2.5,-1.5)(5,1.5)%
\qbezier(1.7,2.5)(0,1.5)(1.7,0.5)%
\qbezier(3.3,2.5)(5,1.5)(3.3,0.5)%
\put(5,1.5){\vector(1,-1){0}}%
\put(5,1.5){\vector(1,1){0}}%
\put(1.7,0.5){\vector(3,-2){0}}%
\put(3.3,0.5){\vector(-3,-2){0}}%
\put(2.5,2.5){\vector(0,-1){2}}%
\put(1.2,1.5){\vector(1,0){1}}%
\put(2.8,1.5){\vector(1,0){1}}%
\end{picture}}

\mcm{\gspecialtwo}{0}{%
\ginitdims{5}{3.4}%
\present{\pregspecialtwo}}

\newcommand{\pregspecialthree}{%
\begin{picture}(5,5.4)(0,-1.2)%
\qbezier(0,1.5)(2.5,6.5)(5,1.5)%
\qbezier(0,1.5)(2.5,-3.5)(5,1.5)%
\qbezier(2,3.5)(1,2.75)(2,2)%
\qbezier(3,3.5)(4,2.75)(3,2)%
\qbezier(2,1)(1,0.25)(2,-0.5)%
\qbezier(3,1)(4,0.25)(3,-0.5)%
\put(0,1.5){\vector(1,0){5}}%
\put(1.5,2.75){\vector(1,0){2}}%
\put(1.5,0.25){\vector(1,0){2}}%
\put(5,1.5){\vector(1,-3){0}}%
\put(5,1.5){\vector(1,3){0}}%
\put(2,2){\vector(1,-1){0}}%
\put(3,2){\vector(-1,-1){0}}%
\put(2,-0.5){\vector(1,-1){0}}%
\put(3,-0.5){\vector(-1,-1){0}}%
\end{picture}}

\mcm{\gspecialthree}{0}{%
\ginitdims{5}{5.4}%
\present{\pregspecialthree}}

\newcommand{\pregonew}[1]{%
\begin{picture}(8,0.4)(0,-0.2)%
\cell{4}{0.2}{b}{#1}%
\put(0,0){\vector(1,0){8}}%
\end{picture}}

\mcm{\gonew}{1}{%
\ginitdims{8}{0.4}%
\abovepic{#1}%
\present{\pregonew{#1}}}

\mcm{\gzersu}{0}{%
\gsinitdims{0}{.6}%
\present{\pregblw{}}}

\mcm{\gonesu}{0}{%
\gsinitdims{5}{0.4}%
\present{\pregone{}}}

\mcm{\gtwosu}{0}{%
\gsinitdims{5}{3.4}%
\present{\pregtwo{}{}{}}}

\mcm{\gthreesu}{0}{%
\gsinitdims{5}{5.4}%
\present{\pregthree{}{}{}{}{}}}

\mcm{\gfoursu}{0}{%
\gsinitdims{5}{8.4}%
\present{\pregfour{}{}{}{}{}{}{}}}

\newcommand{\precone}[1]{%
\begin{picture}(4.2,0.4)(-0.3,-0.2)%
\cell{1.8}{0.2}{b}{#1}%
\put(0,0){\vector(1,0){3.6}}%
\end{picture}}

\mcm{\cone}{1}{%
\cinitdims{4.2}{0.4}%
\abovepic{#1}%
\present{\precone{#1}}}

\mcm{\gfstsu}{0}{%
\gsinitdims{0.5}{0.4}%
\presentr{\pregfst{}}}

\mcm{\glstsu}{0}{%
\gsinitdims{0.5}{0.4}%
\presentl{\preglst{}}}

\newcommand{\prectwodbl}[3]%
{\begin{picture}(4.2,3.4)(-0.1,-0.2)%
\cell{2}{3.2}{b}{#1}%
\cell{2}{-0.2}{t}{#2}%
\cell{2.3}{1.5}{l}{#3}%
\qbezier(0,2)(2,4)(4,2)%
\qbezier(0,1)(2,-1)(4,1)%
\put(4,2){\vector(1,-1){0}}%
\put(4,1){\vector(1,1){0}}%
\put(1.9,2.5){\line(0,-1){1.8}}%
\put(2.1,2.5){\line(0,-1){1.8}}%
\cell{2.01}{0.4}{b}{\vee}%
\end{picture}}

\mcm{\ctwodbl}{3}{%
\cinitdims{4.2}{3.4}%
\abovepic{#1}%
\belowpic{#2}%
\present{\prectwodbl{#1}{#2}{#3}}}

\newcommand{\precthreedbl}[5]{%
\begin{picture}(4.2,5.4)(-0.1,-0.2)%
\cell{2}{5.2}{b}{#1}%
\cell{1}{2.7}{b}{#2}%
\cell{2}{-.2}{t}{#3}%
\cell{2.3}{3.75}{l}{#4}%
\cell{2.3}{1.25}{l}{#5}%
\qbezier(0,3)(2,7)(4,3)%
\qbezier(0,2)(2,-2)(4,2)%
\put(0,2.5){\vector(1,0){4}}%
\put(1.9,4.5){\line(0,-1){1.3}}%
\put(2.1,4.5){\line(0,-1){1.3}}%
\cell{2.01}{2.9}{b}{\vee}%
\put(1.9,2){\line(0,-1){1.3}}%
\put(2.1,2){\line(0,-1){1.3}}%
\cell{2.01}{0.4}{b}{\vee}%
\put(4,3){\vector(1,-3){0}}%
\put(4,2){\vector(1,3){0}}%
\end{picture}}

\mcm{\cthreedbl}{5}{%
\cinitdims{4.2}{5.4}%
\abovepic{#1}%
\belowpic{#3}%
\present{\precthreedbl{#1}{#2}{#3}{#4}{#5}}}

\newcommand{\precthreecelltrp}[5]{%
\begin{picture}(8.2,5)(-4.1,-2.5)%
\cell{0}{2.5}{b}{#1}%
\cell{0}{-2.5}{t}{#2}%
\cell{-1.8}{0}{r}{#3}%
\cell{1.8}{0}{l}{#4}%
\cell{0}{0.3}{b}{#5}%
\qbezier(-4,0.5)(0,4)(4,0.5)%
\qbezier(-4,-0.5)(0,-4)(4,-0.5)%
\qbezier(-0.6,2)(-2.6,0)(-0.6,-2)%
\qbezier(-0.4,2)(-2.4,0)(-0.5,-1.9)%
\cell{-0.6}{-2}{b}{\lrcorner}%
\qbezier(0.4,2)(2.4,0)(0.5,-1.9)%
\qbezier(0.6,2)(2.6,0)(0.6,-2)%
\cell{0.65}{-2}{b}{\llcorner}%
\put(-1,0.15){\line(1,0){1.7}}%
\put(-1,0){\line(1,0){2}}%
\put(-1,-0.15){\line(1,0){1.7}}%
\cell{1.15}{0}{r}{>}%
\put(4,0.5){\vector(1,-1){0}}%
\put(4,-0.5){\vector(1,1){0}}%
\end{picture}}

\mcm{\cthreecelltrp}{5}{%
\cinitdims{8.2}{5}%
\abovepic{#1}%
\belowpic{#2}%
\present{\precthreecelltrp{#1}{#2}{#3}{#4}{#5}}}

\newcommand{\prectwo}[3]%
{\begin{picture}(4.2,3.4)(-0.1,-0.2)%
\cell{2}{3.2}{b}{#1}%
\cell{2}{-0.2}{t}{#2}%
\cell{2.2}{1.5}{l}{#3}%
\qbezier(0,2)(2,4)(4,2)%
\qbezier(0,1)(2,-1)(4,1)%
\put(4,2){\vector(1,-1){0}}%
\put(4,1){\vector(1,1){0}}%
\put(2,2.5){\vector(0,-1){2}}%
\end{picture}}

\mcm{\ctwo}{3}{%
\cinitdims{4.2}{3.4}%
\abovepic{#1}%
\belowpic{#2}%
\present{\prectwo{#1}{#2}{#3}}}

\newcommand{\precthree}[5]{%
\begin{picture}(4.2,5.4)(-0.1,-0.2)%
\cell{2}{5.2}{b}{#1}%
\cell{1}{2.7}{b}{#2}%
\cell{2}{-.2}{t}{#3}%
\cell{2.2}{3.75}{l}{#4}%
\cell{2.2}{1.25}{l}{#5}%
\qbezier(0,3)(2,7)(4,3)%
\qbezier(0,2)(2,-2)(4,2)%
\put(0,2.5){\vector(1,0){4}}%
\put(2,4.5){\vector(0,-1){1.5}}%
\put(2,2){\vector(0,-1){1.5}}%
\put(4,3){\vector(1,-3){0}}%
\put(4,2){\vector(1,3){0}}%
\end{picture}}

\mcm{\cthree}{5}{%
\cinitdims{4.2}{5.4}%
\abovepic{#1}%
\belowpic{#3}%
\present{\precthree{#1}{#2}{#3}{#4}{#5}}}

\newcommand{\prectwoop}[3]%
{\begin{picture}(4.2,3.4)(-0.1,-0.2)%
\cell{2}{3.2}{b}{#1}%
\cell{2}{-0.2}{t}{#2}%
\cell{2.2}{1.5}{l}{#3}%
\qbezier(0,2)(2,4)(4,2)%
\qbezier(0,1)(2,-1)(4,1)%
\put(0,2){\vector(-1,-1){0}}%
\put(0,1){\vector(-1,1){0}}%
\put(2,2.5){\vector(0,-1){2}}%
\end{picture}}

\mcm{\ctwoop}{3}{%
\cinitdims{4.2}{3.4}%
\abovepic{#1}%
\belowpic{#2}%
\present{\prectwoop{#1}{#2}{#3}}}

\newcommand{\prectwopar}[4]{%
\begin{picture}(4.2,3.4)(-0.1,-0.2)%
\cell{2}{3.2}{b}{#1}%
\cell{2}{-0.2}{t}{#2}%
\cell{1.6}{1.5}{r}{#3}%
\cell{2.4}{1.5}{l}{#4}%
\qbezier(0,2)(2,4)(4,2)%
\qbezier(0,1)(2,-1)(4,1)%
\put(4,2){\vector(1,-1){0}}%
\put(4,1){\vector(1,1){0}}%
\put(1.8,2.5){\vector(0,-1){2}}%
\put(2.2,2.5){\vector(0,-1){2}}%
\end{picture}}

\mcm{\ctwopar}{4}{%
\cinitdims{4.2}{3.4}%
\abovepic{#1}%
\belowpic{#2}%
\present{\prectwopar{#1}{#2}{#3}{#4}}}

\newcommand{\precthreein}[5]{%
\begin{picture}(4.2,5.4)(-0.1,-0.2)%
\cell{2}{5.2}{b}{#1}%
\cell{1}{2.7}{b}{#2}%
\cell{2}{-.2}{t}{#3}%
\cell{2.2}{3.75}{l}{#4}%
\cell{2.2}{1.25}{l}{#5}%
\qbezier(0,3)(2,7)(4,3)%
\qbezier(0,2)(2,-2)(4,2)%
\put(0,2.5){\vector(1,0){4}}%
\put(2,4.5){\vector(0,-1){1.5}}%
\put(2,0.5){\vector(0,1){1.5}}%
\put(4,3){\vector(1,-3){0}}%
\put(4,2){\vector(1,3){0}}%
\end{picture}}

\mcm{\cthreein}{5}{%
\cinitdims{4.2}{5.4}%
\abovepic{#1}%
\belowpic{#3}%
\present{\precthreein{#1}{#2}{#3}{#4}{#5}}}

\newcommand{\precthreecell}[5]{%
\begin{picture}(8.2,5)(-4.1,-2.5)%
\cell{0}{2.5}{b}{#1}%
\cell{0}{-2.5}{t}{#2}%
\cell{-1.7}{0}{r}{#3}%
\cell{1.7}{0}{l}{#4}%
\cell{0}{0.2}{b}{#5}%
\qbezier(-4,0.5)(0,4)(4,0.5)%
\qbezier(-4,-0.5)(0,-4)(4,-0.5)%
\qbezier(-0.5,2)(-2.5,0)(-0.5,-2)%
\qbezier(0.5,2)(2.5,0)(0.5,-2)%
\put(-1,0){\vector(1,0){2}}%
\put(4,0.5){\vector(1,-1){0}}%
\put(4,-0.5){\vector(1,1){0}}%
\put(-0.5,-2){\vector(1,-1){0}}%
\put(0.5,-2){\vector(-1,-1){0}}%
\end{picture}}

\mcm{\cthreecell}{5}{%
\cinitdims{8.2}{5}%
\abovepic{#1}%
\belowpic{#2}%
\present{\precthreecell{#1}{#2}{#3}{#4}{#5}}}

\newcommand{\precthreecellpar}[6]{%
\begin{picture}(8.2,5)(-4.1,-2.5)%
\cell{0}{2.5}{b}{#1}%
\cell{0}{-2.5}{t}{#2}%
\cell{-1.7}{0}{r}{#3}%
\cell{1.7}{0}{l}{#4}%
\cell{0}{0.4}{b}{#5}%
\cell{0}{-0.4}{t}{#6}%
\qbezier(-4,0.5)(0,4)(4,0.5)%
\qbezier(-4,-0.5)(0,-4)(4,-0.5)%
\qbezier(-0.5,2)(-2.5,0)(-0.5,-2)%
\qbezier(0.5,2)(2.5,0)(0.5,-2)%
\put(-1,0.2){\vector(1,0){2}}%
\put(-1,-0.2){\vector(1,0){2}}%
\put(4,0.5){\vector(1,-1){0}}%
\put(4,-0.5){\vector(1,1){0}}%
\put(-0.5,-2){\vector(1,-1){0}}%
\put(0.5,-2){\vector(-1,-1){0}}%
\end{picture}}

\mcm{\cthreecellpar}{6}{%
\cinitdims{8.2}{5}%
\abovepic{#1}%
\belowpic{#2}%
\present{\precthreecellpar{#1}{#2}{#3}{#4}{#5}{#6}}}

\newcommand{\prectwov}[5]{%
\begin{picture}(3.4,4.2)(0.8,0.9)%
\cell{2.5}{5.1}{b}{#1}%
\cell{2.5}{0.9}{t}{#2}%
\cell{0.8}{3}{r}{#3}%
\cell{4.2}{3}{l}{#4}%
\cell{2.5}{3.2}{b}{#5}%
\qbezier(2,5)(0,3)(2,1)%
\qbezier(3,5)(5,3)(3,1)%
\put(2,1){\vector(1,-1){0}}%
\put(3,1){\vector(-1,-1){0}}%
\put(1.5,3){\vector(1,0){2}}%
\end{picture}}

\mcm{\ctwov}{5}{%
\cinitdims{3.4}{4.2}%
\abovepic{#1}%
\belowpic{#2}%
\sidespic{#3}%
\sidespic{#4}%
\present{\prectwov{#1}{#2}{#3}{#4}{#5}}}

\newcommand{\precthreecellv}[7]{%
\begin{picture}(5,8.2)(0.5,-1.6)%
\cell{3}{6.6}{b}{#1}%
\cell{3}{-1.6}{t}{#2}%
\cell{0.5}{2.5}{r}{#3}%
\cell{5.5}{2.5}{l}{#4}%
\cell{3}{4.2}{b}{#5}%
\cell{3}{0.8}{t}{#6}%
\cell{3.2}{2.5}{l}{#7}%
\qbezier(3.5,6.5)(7,2.5)(3.5,-1.5)%
\qbezier(2.5,6.5)(-1,2.5)(2.5,-1.5)%
\put(2.5,-1.5){\vector(1,-1){0}}%
\put(3.5,-1.5){\vector(-1,-1){0}}%
\qbezier(1,3)(3,5)(5,3)%
\qbezier(1,2)(3,0)(5,2)%
\put(5,3){\vector(1,-1){0}}%
\put(5,2){\vector(1,1){0}}%
\put(3,3.5){\vector(0,-1){2}}%
\end{picture}}

\mcm{\cthreecellv}{7}{%
\cinitdims{5}{8.2}%
\abovepic{#1}%
\belowpic{#2}%
\sidespic{#3}%
\sidespic{#4}%
\present{\precthreecellv{#1}{#2}{#3}{#4}{#5}{#6}{#7}}}

\newcommand{\pretopez}[2]{%
\begin{picture}(2.6,2.3)(-1.3,-2.2)%
\cell{0}{-2.2}{t}{#1}%
\cell{0}{-1.2}{c}{#2}%
\qbezier(0,0)(-2,-2)(0,-2)%
\qbezier(0,0)(2,-2)(0,-2)%
\put(0,0){\vector(-1,1){0}}%
\end{picture}}

\mcm{\topez}{2}{%
\ginitdims{2.6}{2.3}%
\belowpic{#1}%
\present{\pretopez{#1}{#2}}}

\newcommand{\pretopea}[3]{%
\begin{picture}(4,1.9)(-2,-0,2)%
\cell{0}{1.7}{b}{#1}%
\cell{0}{-0.2}{t}{#2}%
\cell{0}{0.7}{c}{#3}%
\qbezier(-2,0)(0,3)(2,0)%
\put(-2,0){\vector(1,0){4}}%
\put(2,0){\vector(2,-3){0}}%
\end{picture}}

\mcm{\topea}{3}{%
\ginitdims{4}{1.9}%
\abovepic{#1}%
\belowpic{#2}%
\present{\pretopea{#1}{#2}{#3}}}

\newcommand{\pretopeb}[4]{%
\begin{picture}(4,2.2)(-2,-0.2)%
\cell{-1.1}{1}{br}{#1}%
\cell{1.1}{1}{bl}{#2}%
\cell{0}{-0.2}{t}{#3}%
\cell{0}{0.8}{c}{#4}%
\put(-2,0){\vector(1,1){2}}%
\put(0,2){\vector(1,-1){2}}%
\put(-2,0){\vector(1,0){4}}%
\end{picture}}

\mcm{\topeb}{4}{%
\ginitdims{4}{2.2}%
\belowpic{#3}%
\present{\pretopeb{#1}{#2}{#3}{#4}}}

\newcommand{\pretopec}[5]{%
\begin{picture}(4,2.2)(-2,-0.2)%
\cell{-1.8}{1}{br}{#1}%
\cell{0}{2.2}{b}{#2}%
\cell{1.8}{1}{bl}{#3}%
\cell{0}{-0.2}{t}{#4}%
\cell{0}{0.8}{c}{#5}%
\put(-2,0){\vector(1,2){1}}%
\put(-1,2){\vector(1,0){2}}%
\put(1,2){\vector(1,-2){1}}%
\put(-2,0){\vector(1,0){4}}%
\end{picture}}

\mcm{\topec}{5}{%
\ginitdims{4}{2.2}%
\sidespic{#1}%
\abovepic{#2}%
\sidespic{#3}%
\belowpic{#4}%
\present{\pretopec{#1}{#2}{#3}{#4}{#5}}}

\newcommand{\pretoped}[6]{%
\begin{picture}(4,2.5)(-2,-0.2)%
\cell{-2}{0.6}{br}{#1}%
\cell{-0.7}{2.2}{br}{#2}%
\cell{0.7}{2.2}{bl}{#3}%
\cell{2}{0.6}{bl}{#4}%
\cell{0}{-0.2}{t}{#5}%
\cell{0}{0.8}{c}{#6}%
\put(-2,0){\vector(1,3){0.5}}%
\put(-1.5,1.5){\vector(3,2){1.5}}%
\put(0,2.5){\vector(3,-2){1.5}}%
\put(1.5,1.5){\vector(1,-3){0.5}}%
\put(-2,0){\vector(1,0){4}}%
\end{picture}}

\mcm{\toped}{6}{%
\ginitdims{4}{2.5}%
\sidespic{#1}%
\abovepic{#2}%
\abovepic{#3}%
\sidespic{#4}%
\belowpic{#5}%
\present{\pretoped{#1}{#2}{#3}{#4}{#5}{#6}}}

\newcommand{\pretopeq}[5]{%
\begin{picture}(4,2.5)(-2,-0.2)%
\cell{-2}{0.6}{br}{#1}%
\cell{-1}{2.2}{br}{#2}%
\cell{2}{0.6}{bl}{#3}%
\cell{0}{-0.2}{t}{#4}%
\cell{0}{0.8}{c}{#5}%
\put(-2,0){\vector(1,3){0.5}}%
\put(-1.5,1.5){\vector(1,1){1}}%
\cell{0.9}{2.3}{c}{\ddots}
\put(1.5,1.5){\vector(1,-3){0.5}}%
\put(-2,0){\vector(1,0){4}}%
\end{picture}}

\mcm{\topeq}{5}{%
\ginitdims{4}{2.5}%
\sidespic{#1}%
\abovepic{#2}%
\sidespic{#3}%
\belowpic{#4}%
\present{\pretopeq{#1}{#2}{#3}{#4}{#5}}}

\newcommand{\pretopebase}[1]{%
\begin{picture}(4,0.4)(0,-0.2)%
\cell{2}{0.2}{b}{#1}%
\put(0,0){\vector(1,0){4}}%
\end{picture}}

\mcm{\topebase}{1}{%
\ginitdims{4}{0.4}%
\abovepic{#1}%
\present{\pretopebase{#1}}}

\newcommand{\pretopezs}[2]{%
\begin{picture}(2.6,2.3)(-1.3,-2.2)%
\cell{0}{-2.2}{t}{#1}%
\cell{0}{-1.2}{c}{#2}%
\qbezier(0,0)(-2,-2)(0,-2)%
\qbezier(0,0)(2,-2)(0,-2)%
\end{picture}}

\mcm{\topezs}{2}{%
\ginitdims{2.6}{2.3}%
\belowpic{#1}%
\present{\pretopezs{#1}{#2}}}

\newcommand{\pretopeas}[3]{%
\begin{picture}(4,1.9)(-2,-0,2)%
\cell{0}{1.7}{b}{#1}%
\cell{0}{-0.2}{t}{#2}%
\cell{0}{0.7}{c}{#3}%
\qbezier(-2,0)(0,3)(2,0)%
\put(-2,0){\line(1,0){4}}%
\end{picture}}

\mcm{\topeas}{3}{%
\ginitdims{4}{1.9}%
\abovepic{#1}%
\belowpic{#2}%
\present{\pretopeas{#1}{#2}{#3}}}

\newcommand{\pretopebs}[4]{%
\begin{picture}(4,2.2)(-2,-0.2)%
\cell{-1.1}{1}{br}{#1}%
\cell{1.1}{1}{bl}{#2}%
\cell{0}{-0.2}{t}{#3}%
\cell{0}{0.8}{c}{#4}%
\put(-2,0){\line(1,1){2}}%
\put(0,2){\line(1,-1){2}}%
\put(-2,0){\line(1,0){4}}%
\end{picture}}

\mcm{\topebs}{4}{%
\ginitdims{4}{2.2}%
\belowpic{#3}%
\present{\pretopebs{#1}{#2}{#3}{#4}}}

\newcommand{\pretopecs}[5]{%
\begin{picture}(4,2.2)(-2,-0.2)%
\cell{-1.8}{1}{br}{#1}%
\cell{0}{2.2}{b}{#2}%
\cell{1.8}{1}{bl}{#3}%
\cell{0}{-0.2}{t}{#4}%
\cell{0}{0.8}{c}{#5}%
\put(-2,0){\line(1,2){1}}%
\put(-1,2){\line(1,0){2}}%
\put(1,2){\line(1,-2){1}}%
\put(-2,0){\line(1,0){4}}%
\end{picture}}

\mcm{\topecs}{5}{%
\ginitdims{4}{2.2}%
\sidespic{#1}%
\abovepic{#2}%
\sidespic{#3}%
\belowpic{#4}%
\present{\pretopecs{#1}{#2}{#3}{#4}{#5}}}

\newcommand{\pretopeds}[6]{%
\begin{picture}(4,2.5)(-2,-0.2)%
\cell{-2}{0.6}{br}{#1}%
\cell{-0.7}{2.2}{br}{#2}%
\cell{0.7}{2.2}{bl}{#3}%
\cell{2}{0.6}{bl}{#4}%
\cell{0}{-0.2}{t}{#5}%
\cell{0}{0.8}{c}{#6}%
\put(-2,0){\line(1,3){0.5}}%
\put(-1.5,1.5){\line(3,2){1.5}}%
\put(0,2.5){\line(3,-2){1.5}}%
\put(1.5,1.5){\line(1,-3){0.5}}%
\put(-2,0){\line(1,0){4}}%
\end{picture}}

\mcm{\topeds}{6}{%
\ginitdims{4}{2.5}%
\sidespic{#1}%
\abovepic{#2}%
\abovepic{#3}%
\sidespic{#4}%
\belowpic{#5}%
\present{\pretopeds{#1}{#2}{#3}{#4}{#5}{#6}}}

\newcommand{\pretopeqs}[5]{%
\begin{picture}(4,2.5)(-2,-0.2)%
\cell{-2}{0.6}{br}{#1}%
\cell{-1}{2.2}{br}{#2}%
\cell{2}{0.6}{bl}{#3}%
\cell{0}{-0.2}{t}{#4}%
\cell{0}{0.8}{c}{#5}%
\put(-2,0){\line(1,3){0.5}}%
\put(-1.5,1.5){\line(1,1){1}}%
\cell{0.9}{2.3}{c}{\ddots}
\put(1.5,1.5){\line(1,-3){0.5}}%
\put(-2,0){\line(1,0){4}}%
\end{picture}}

\mcm{\topeqs}{5}{%
\ginitdims{4}{2.5}%
\sidespic{#1}%
\abovepic{#2}%
\sidespic{#3}%
\belowpic{#4}%
\present{\pretopeqs{#1}{#2}{#3}{#4}{#5}}}

\newcommand{\pretopebases}[1]{%
\begin{picture}(4,0.4)(0,-0.2)%
\cell{2}{0.2}{b}{#1}%
\put(0,0){\line(1,0){4}}%
\end{picture}}

\mcm{\topebases}{1}{%
\ginitdims{4}{0.4}%
\abovepic{#1}%
\present{\pretopebases{#1}}}

\newcommand{\pregdots}[6]{%
\begin{picture}(5,8.4)(0,-2.7)%
\cell{2.5}{5.7}{b}{#1}%
\cell{1.5}{2.8}{b}{#2}%
\cell{1.5}{0.2}{t}{#3}%
\cell{2.5}{-2.7}{t}{#4}%
\cell{2.7}{4.25}{l}{#5}%
\cell{2.7}{-1.25}{l}{#6}%
\qbezier(0,1.5)(2.5,9.5)(5,1.5)%
\qbezier(0,1.5)(2.5,4)(5,1.5)%
\qbezier(0,1.5)(2.5,-1)(5,1.5)%
\qbezier(0,1.5)(2.5,-6.5)(5,1.5)%
\put(2.5,5.25){\vector(0,-1){2}}%
\put(2.5,-0.25){\vector(0,-1){2}}%
\cell{2.5}{1.7}{c}{\vdots}%
\put(5,1.5){\vector(1,-4){0}}%
\put(5,1.5){\vector(4,-3){0}}%
\put(5,1.5){\vector(4,3){0}}%
\put(5,1.5){\vector(1,4){0}}%
\end{picture}}

\mcm{\gdots}{6}{%
\ginitdims{5}{8.4}%
\abovepic{#1}%
\belowpic{#4}%
\present{\pregdots{#1}{#2}{#3}{#4}{#5}{#6}}}

\newlength{\volt}
\makeatletter

\def\diagram{\m@th\leftwidth=\z@ \rightwidth=\z@ \topheight=\z@
\botheight=\z@ \setbox\@picbox\hbox\bgroup}

\def\enddiagram{\egroup\wd\@picbox\rightwidth\unitlength
\ht\@picbox\topheight\unitlength \dp\@picbox\botheight\unitlength
\hskip\leftwidth\unitlength\box\@picbox}

\def\bfig{\begin{diagram}}
\def\efig{\end{diagram}}
\newcount\wideness \newcount\leftwidth \newcount\rightwidth
\newcount\highness \newcount\topheight \newcount\botheight

\def\ratchet#1#2{\ifnum#1<#2 \global #1=#2 \fi}

\def\putbox(#1,#2)#3{%
\horsize{\wideness}{#3} \divide\wideness by 2 {\advance\wideness
by #1 \ratchet{\rightwidth}{\wideness}} {\advance\wideness by -#1
\ratchet{\leftwidth}{\wideness}} \vertsize{\highness}{#3}
\divide\highness by 2 {\advance\highness by #2
\ratchet{\topheight}{\highness}} {\advance\highness by -#2
\ratchet{\botheight}{\highness}} \put(#1,#2){\makebox(0,0){$#3$}}}

\def\putlbox(#1,#2)#3{%
\horsize{\wideness}{#3} {\advance\wideness by #1
\ratchet{\rightwidth}{\wideness}} {\ratchet{\leftwidth}{-#1}}
\vertsize{\highness}{#3} \divide\highness by 2 {\advance\highness
by #2 \ratchet{\topheight}{\highness}} {\advance\highness by -#2
\ratchet{\botheight}{\highness}}
\put(#1,#2){\makebox(0,0)[l]{$#3$}}}

\def\putrbox(#1,#2)#3{%
\horsize{\wideness}{#3} {\ratchet{\rightwidth}{#1}}
{\advance\wideness by -#1 \ratchet{\leftwidth}{\wideness}}
\vertsize{\highness}{#3} \divide\highness by 2 {\advance\highness
by #2 \ratchet{\topheight}{\highness}} {\advance\highness by -#2
\ratchet{\botheight}{\highness}}
\put(#1,#2){\makebox(0,0)[r]{$#3$}}}

\def\adjust[#1]{} 

\newcount \coefa
\newcount \coefb
\newcount \coefc
\newcount\tempcounta
\newcount\tempcountb
\newcount\tempcountc
\newcount\tempcountd
\newcount\xext
\newcount\yext
\newcount\xoff
\newcount\yoff
\newcount\gap%
\newcount\arrowtypea
\newcount\arrowtypeb
\newcount\arrowtypec
\newcount\arrowtyped
\newcount\arrowtypee
\newcount\height
\newcount\width
\newcount\xpos
\newcount\ypos
\newcount\run
\newcount\rise
\newcount\arrowlength
\newcount\halflength
\newcount\arrowtype
\newdimen\tempdimen
\newdimen\xlen
\newdimen\ylen
\newsavebox{\tempboxa}%
\newsavebox{\tempboxb}%
\newsavebox{\tempboxc}%

\newdimen\w@dth

\def\setw@dth#1#2{\setbox\z@\hbox{\m@th$#1$}\w@dth=\wd\z@
\setbox\@ne\hbox{\m@th$#2$}\ifnum\w@dth<\wd\@ne \w@dth=\wd\@ne \fi
\advance\w@dth by 1.2em}

\def\t@^#1_#2{\allowbreak\def\n@one{#1}\def\n@two{#2}\mathrel
{\setw@dth{#1}{#2} \mathop{\hbox to
\w@dth{\rightarrowfill}}\limits \ifx\n@one\empty\else
^{\box\z@}\fi \ifx\n@two\empty\else _{\box\@ne}\fi}}
\def\t@@^#1{\@ifnextchar_{\t@^{#1}}{\t@^{#1}_{}}}
\def\to{\@ifnextchar^{\t@@}{\t@@^{}}}

\def\t@left^#1_#2{\def\n@one{#1}\def\n@two{#2}\mathrel{\setw@dth{#1}{#2}
\mathop{\hbox to \w@dth{\leftarrowfill}}\limits
\ifx\n@one\empty\else ^{\box\z@}\fi \ifx\n@two\empty\else
_{\box\@ne}\fi}}
\def\t@@left^#1{\@ifnextchar_{\t@left^{#1}}{\t@left^{#1}_{}}}
\def\toleft{\@ifnextchar^{\t@@left}{\t@@left^{}}}

\def\two@^#1_#2{\allowbreak
\def\n@one{#1}\def\n@two{#2}\mathrel{\setw@dth{#1}{#2}
\mathop{\vcenter{\lineskip\z@\baselineskip\z@
                 \hbox to \w@dth{\rightarrowfill}%
                 \hbox to \w@dth{\rightarrowfill}}%
       }\limits
\ifx\n@one\empty\else ^{\box\z@}\fi \ifx\n@two\empty\else
_{\box\@ne}\fi}}
\def\tw@@^#1{\@ifnextchar _{\two@^{#1}}{\two@^{#1}_{}}}
\def\two{\@ifnextchar ^{\tw@@}{\tw@@^{}}}

\def\tofr@^#1_#2{\def\n@one{#1}\def\n@two{#2}\mathrel{\setw@dth{#1}{#2}
\mathop{\vcenter{\hbox to \w@dth{\rightarrowfill}\kern-1.7ex
                 \hbox to \w@dth{\leftarrowfill}}%
       }\limits
\ifx\n@one\empty\else ^{\box\z@}\fi \ifx\n@two\empty\else
_{\box\@ne}\fi}}
\def\t@fr@^#1{\@ifnextchar_ {\tofr@^{#1}}{\tofr@^{#1}_{}}}
\def\tofro{\@ifnextchar^ {\t@fr@}{\t@fr@^{}}}

\def\mon{\mathop{\m@th\hbox to
      14.6\P@{\lasyb\char'51\hskip-2.1\P@$\arrext$\hss
$\mathord\rightarrow$}}\limits} 
\def\leftmono{\mathrel{\m@th\hbox to
14.6\P@{$\mathord\leftarrow$\hss$\arrext$\hskip-2.1\P@\lasyb\char'50%
}}\limits} 
\mathchardef\arrext="0200       

\def\settypes(#1,#2,#3){\arrowtypea#1 \arrowtypeb#2 \arrowtypec#3}
\def\settoheight#1#2{\setbox\@tempboxa\hbox{#2}#1\ht\@tempboxa\relax}%
\def\settodepth#1#2{\setbox\@tempboxa\hbox{#2}#1\dp\@tempboxa\relax}%
\def\settokens`#1`#2`#3`#4`{%
     \def\tokena{#1}\def\tokenb{#2}\def\tokenc{#3}\def\tokend{#4}}
\def\setsqparms[#1`#2`#3`#4;#5`#6]{%
\arrowtypea #1 \arrowtypeb #2 \arrowtypec #3 \arrowtyped #4
\width #5 \height #6 }
\def\setpos(#1,#2){\xpos=#1 \ypos#2}

\def\settriparms[#1`#2`#3;#4]{\settripairparms[#1`#2`#3`1`1;#4]}%

\def\settripairparms[#1`#2`#3`#4`#5;#6]{%
\arrowtypea #1 \arrowtypeb #2 \arrowtypec #3 \arrowtyped #4
\arrowtypee #5 \width #6 \height #6 }

\def\resetparms{\settripairparms[1`1`1`1`1;500]\width 500}

\resetparms

\def\mvector(#1,#2)#3{
\put(0,0){\vector(#1,#2){#3}}%
\put(0,0){\vector(#1,#2){26}}%
}
\def\evector(#1,#2)#3{{
\arrowlength #3
\put(0,0){\vector(#1,#2){\arrowlength}}%
\advance \arrowlength by-30
\put(0,0){\vector(#1,#2){\arrowlength}}%
}}

\def\horsize#1#2{%
\settowidth{\tempdimen}{$#2$}%
#1=\tempdimen \divide #1 by\unitlength }

\def\vertsize#1#2{%
\settoheight{\tempdimen}{$#2$}%
#1=\tempdimen
\settodepth{\tempdimen}{$#2$}%
\advance #1 by\tempdimen \divide #1 by\unitlength }

\def\putvector(#1,#2)(#3,#4)#5#6{{%
\ifnum3<\arrowtype \putdashvector(#1,#2)(#3,#4)#5\arrowtype \else
\ifnum\arrowtype<-3 \putdashvector(#1,#2)(#3,#4)#5\arrowtype \else
\xpos=#1 \ypos=#2 \run=#3 \rise=#4 \arrowlength=#5 \ifnum
\arrowtype<0
    \ifnum \run=0
        \advance \ypos by-\arrowlength
    \else
        \tempcounta \arrowlength
        \multiply \tempcounta by\rise
        \divide \tempcounta by\run
        \ifnum\run>0
            \advance \xpos by\arrowlength
            \advance \ypos by\tempcounta
        \else
            \advance \xpos by-\arrowlength
            \advance \ypos by-\tempcounta
        \fi
    \fi
    \multiply \arrowtype by-1
    \multiply \rise by-1
    \multiply \run by-1
\fi \ifcase \arrowtype
\or \put(\xpos,\ypos){\vector(\run,\rise){\arrowlength}}%
\or \put(\xpos,\ypos){\mvector(\run,\rise)\arrowlength}%
\or \put(\xpos,\ypos){\evector(\run,\rise){\arrowlength}}%
\fi\fi\fi }}

\def\putsplitvector(#1,#2)#3#4{
\xpos #1 \ypos #2 \arrowtype #4 \halflength #3 \arrowlength #3
\gap 140 \advance \halflength by-\gap \divide \halflength by2
\ifnum\arrowtype>0
   \ifcase \arrowtype
   \or \put(\xpos,\ypos){\line(0,-1){\halflength}}%
       \advance\ypos by-\halflength
       \advance\ypos by-\gap
       \put(\xpos,\ypos){\vector(0,-1){\halflength}}%
   \or \put(\xpos,\ypos){\line(0,-1)\halflength}%
       \put(\xpos,\ypos){\vector(0,-1)3}%
       \advance\ypos by-\halflength
       \advance\ypos by-\gap
       \put(\xpos,\ypos){\vector(0,-1){\halflength}}%
   \or \put(\xpos,\ypos){\line(0,-1)\halflength}%
       \advance\ypos by-\halflength
       \advance\ypos by-\gap
       \put(\xpos,\ypos){\evector(0,-1){\halflength}}%
   \fi
\else \arrowtype=-\arrowtype
   \ifcase\arrowtype
   \or \advance \ypos by-\arrowlength
       \put(\xpos,\ypos){\line(0,1){\halflength}}%
       \advance\ypos by\halflength
       \advance\ypos by\gap
       \put(\xpos,\ypos){\vector(0,1){\halflength}}%
   \or \advance \ypos by-\arrowlength
       \put(\xpos,\ypos){\line(0,1)\halflength}%
       \put(\xpos,\ypos){\vector(0,1)3}%
       \advance\ypos by\halflength
       \advance\ypos by\gap
       \put(\xpos,\ypos){\vector(0,1){\halflength}}%
   \or \advance \ypos by-\arrowlength
       \put(\xpos,\ypos){\line(0,1)\halflength}%
       \advance\ypos by\halflength
       \advance\ypos by\gap
       \put(\xpos,\ypos){\evector(0,1){\halflength}}%
   \fi
\fi }

\def\putmorphism(#1)(#2,#3)[#4`#5`#6]#7#8#9{{%
\run #2 \rise #3 \ifnum\rise=0
  \puthmorphism(#1)[#4`#5`#6]{#7}{#8}#9%
\else\ifnum\run=0
  \putvmorphism(#1)[#4`#5`#6]{#7}{#8}#9%
\else
\setpos(#1)%
\arrowlength #7 \arrowtype #8 \ifnum\run=0 \else\ifnum\rise=0
\else \ifnum\run>0
    \coefa=1
\else
   \coefa=-1
\fi \ifnum\arrowtype>0
   \coefb=0
   \coefc=-1
\else
   \coefb=\coefa
   \coefc=1
   \arrowtype=-\arrowtype
\fi \width=2 \multiply \width by\run \divide \width by\rise
\ifnum \width<0  \width=-\width\fi \advance\width by60 \if l#9
\width=-\width\fi
\putbox(\xpos,\ypos){#4}
{\multiply \coefa by\arrowlength
\advance\xpos by\coefa \multiply \coefa by\rise \divide \coefa
by\run \advance \ypos by\coefa
\putbox(\xpos,\ypos){#5} }%
{\multiply \coefa by\arrowlength
\divide \coefa by2 \advance \xpos by\coefa \advance \xpos by\width
\multiply \coefa by\rise \divide \coefa by\run \advance \ypos
by\coefa
\if l#9%
   \putrbox(\xpos,\ypos){#6}%
\else\if r#9%
   \putlbox(\xpos,\ypos){#6}%
\fi\fi }%
{\multiply \rise by-\coefc
\multiply \run by-\coefc \multiply \coefb by\arrowlength \advance
\xpos by\coefb \multiply \coefb by\rise \divide \coefb by\run
\advance \ypos by\coefb \multiply \coefc by70 \advance \ypos
by\coefc \multiply \coefc by\run \divide \coefc by\rise \advance
\xpos by\coefc \multiply \coefa by140 \multiply \coefa by\run
\divide \coefa by\rise \advance \arrowlength by\coefa
\ifcase\arrowtype
\or \put(\xpos,\ypos){\vector(\run,\rise){\arrowlength}}%
\or \put(\xpos,\ypos){\mvector(\run,\rise){\arrowlength}}%
\or \put(\xpos,\ypos){\evector(\run,\rise){\arrowlength}}%
\fi}\fi\fi\fi\fi}}

\newcount\numbdashes \newcount\lengthdash \newcount\increment

\def\howmanydashes{
\numbdashes=\arrowlength \lengthdash=40 \divide\numbdashes by
\lengthdash \lengthdash=\arrowlength \divide\lengthdash by
\numbdashes
\increment=\lengthdash \multiply\lengthdash by 3
\divide\lengthdash by 5 }

\def\putdashvector(#1)(#2,#3)#4#5{%
\ifnum#3=0 \putdashhvector(#1){#4}#5 \else \ifnum#2=0
\putdashvvector(#1){#4}#5\fi\fi}

\def\putdashhvector(#1,#2)#3#4{{%
\arrowlength=#3 \howmanydashes
\multiput(#1,#2)(\increment,0){\numbdashes}%
{\vrule height .4pt width \lengthdash\unitlength} \arrowtype=#4
\xpos=#1 \ifnum\arrowtype<0 \advance\arrowtype by 7 \fi
\ifcase\arrowtype \or \advance\xpos by 10
    \put(\xpos,#2){\vector(-1,0){\lengthdash}}
    \advance\xpos by 40
    \put(\xpos,#2){\vector(-1,0){\lengthdash}}
\or \advance \xpos by 10
    \put(\xpos,#2){\vector(-1,0){\lengthdash}}
    \advance\xpos by  \arrowlength
    \advance\xpos by  -50
    \put(\xpos,#2){\vector(-1,0){\lengthdash}}
\or \advance\xpos by 10
    \put(\xpos,#2){\vector(-1,0){\lengthdash}}
\or \advance\xpos by \arrowlength
    \advance\xpos by -\lengthdash
    \put(\xpos,#2){\vector(1,0){\lengthdash}}
\or {\advance\xpos by 10
    \put(\xpos,#2){\vector(1,0){\lengthdash}}}
    \advance\xpos by \arrowlength
    \advance\xpos by -\lengthdash
    \put(\xpos,#2){\vector(1,0){\lengthdash}}
\or \advance\xpos by \arrowlength
    \advance\xpos by -\lengthdash
    \put(\xpos,#2){\vector(1,0){\lengthdash}}
    \advance\xpos by -40
    \put(\xpos,#2){\vector(1,0){\lengthdash}}
   \fi
}}

\def\putdashvvector(#1,#2)#3#4{{%
\arrowlength=#3 \howmanydashes \ypos=#2 \advance\ypos by
-\arrowlength
\multiput(#1,#2)(0,\increment){\numbdashes}%
    {\vrule width .4pt height \lengthdash\unitlength}
\arrowtype=#4 \ypos=#2 \ifnum\arrowtype<0 \advance\arrowtype by 7
\fi \ifcase\arrowtype \or \advance\ypos by \arrowlength
\advance\ypos by -40
    \put(#1,\ypos){\vector(0,1){\lengthdash}}
    \advance\ypos by -40
    \put(#1,\ypos){\vector(0,1){\lengthdash}}
\or \advance\ypos by 10
    \put(#1,\ypos){\vector(0,1){\lengthdash}}
    \advance\ypos by \arrowlength \advance\ypos by -40
    \put(#1,\ypos){\vector(0,1){\lengthdash}}
\or \advance\ypos by \arrowlength \advance\ypos by -40
    \put(#1,\ypos){\vector(0,1){\lengthdash}}
\or \advance\ypos by 10
    \put(#1,\ypos){\vector(0,-1){\lengthdash}}
\or \advance\ypos by 10
    \put(#1,\ypos){\vector(0,-1){\lengthdash}}
    \advance\ypos by \arrowlength \advance\ypos by -40
    \put(#1,\ypos){\vector(0,-1){\lengthdash}}
\or \advance\ypos by 10
    \put(#1,\ypos){\vector(0,-1){\lengthdash}}
    \advance\ypos by 40
    \put(#1,\ypos){\vector(0,-1){\lengthdash}}
\fi }}

\def\puthmorphism(#1,#2)[#3`#4`#5]#6#7#8{{%
\xpos #1 \ypos #2 \width #6 \arrowlength #6 \arrowtype=#7
\putbox(\xpos,\ypos){#3\vphantom{#4}}%
{\advance \xpos by\arrowlength
\putbox(\xpos,\ypos){\vphantom{#3}#4}}%
\horsize{\tempcounta}{#3}%
\horsize{\tempcountb}{#4}%
\divide \tempcounta by2 \divide \tempcountb by2 \advance
\tempcounta by30 \advance \tempcountb by30 \advance \xpos
by\tempcounta \advance \arrowlength by-\tempcounta \advance
\arrowlength by-\tempcountb
\putvector(\xpos,\ypos)(1,0)\arrowlength\arrowtype \divide
\arrowlength by2 \advance \xpos by\arrowlength
\vertsize{\tempcounta}{#5}%
\divide\tempcounta by2 \advance \tempcounta by20
\if a#8 %
   \advance \ypos by\tempcounta
   \putbox(\xpos,\ypos){#5}%
\else
   \advance \ypos by-\tempcounta
   \putbox(\xpos,\ypos){#5}%
\fi}}

\def\putvmorphism(#1,#2)[#3`#4`#5]#6#7#8{{%
\xpos #1 \ypos #2 \arrowlength #6 \arrowtype #7
\settowidth{\xlen}{$#5$}%
\putbox(\xpos,\ypos){#3}%
{\advance \ypos by-\arrowlength
\putbox(\xpos,\ypos){#4}}%
{\advance\arrowlength by-140 \advance \ypos by-70 \ifdim\xlen>0pt
   \if m#8%
      \putsplitvector(\xpos,\ypos)\arrowlength\arrowtype
   \else
   \putvector(\xpos,\ypos)(0,-1)\arrowlength\arrowtype
   \fi
\else
   \putvector(\xpos,\ypos)(0,-1)\arrowlength\arrowtype
\fi}%
\ifdim\xlen>0pt
   \divide \arrowlength by2
   \advance\ypos by-\arrowlength
   \if l#8%
      \advance \xpos by-40
      \putrbox(\xpos,\ypos){#5}%
   \else\if r#8%
      \advance \xpos by40
      \putlbox(\xpos,\ypos){#5}%
   \else
      \putbox(\xpos,\ypos){#5}%
   \fi\fi
\fi }}

\def\putsquarep<#1>(#2)[#3;#4`#5`#6`#7]{{%
\setsqparms[#1]%
\setpos(#2)%
\settokens`#3`%
\puthmorphism(\xpos,\ypos)[\tokenc`\tokend`{#7}]{\width}{\arrowtyped}b%
\advance\ypos by \height
\puthmorphism(\xpos,\ypos)[\tokena`\tokenb`{#4}]{\width}{\arrowtypea}a%
\putvmorphism(\xpos,\ypos)[``{#5}]{\height}{\arrowtypeb}l%
\advance\xpos by \width
\putvmorphism(\xpos,\ypos)[``{#6}]{\height}{\arrowtypec}r%
}}

\def\putsquare{\@ifnextchar <{\putsquarep}{\putsquarep%
   <\arrowtypea`\arrowtypeb`\arrowtypec`\arrowtyped;\width`\height>}}
\def\square{\@ifnextchar< {\squarep}{\squarep
   <\arrowtypea`\arrowtypeb`\arrowtypec`\arrowtyped;\width`\height>}}
\def\squarep<#1>[#2`#3`#4`#5;#6`#7`#8`#9]{{
\setsqparms[#1]
\diagram
\putsquarep<\arrowtypea`\arrowtypeb`\arrowtypec`
\arrowtyped;\width`\height>
(0,0)[#2`#3`#4`{#5};#6`#7`#8`{#9}]
\enddiagram
}}                                                 
\def\putptrianglep<#1>(#2,#3)[#4`#5`#6;#7`#8`#9]{{%
\settriparms[#1]%
\xpos=#2 \ypos=#3 \advance\ypos by \height
\puthmorphism(\xpos,\ypos)[#4`#5`{#7}]{\height}{\arrowtypea}a%
\putvmorphism(\xpos,\ypos)[`#6`{#8}]{\height}{\arrowtypeb}l%
\advance\xpos by\height
\putmorphism(\xpos,\ypos)(-1,-1)[``{#9}]{\height}{\arrowtypec}r%
}}

\def\putptriangle{\@ifnextchar <{\putptrianglep}{\putptrianglep
   <\arrowtypea`\arrowtypeb`\arrowtypec;\height>}}
\def\ptriangle{\@ifnextchar <{\ptrianglep}{\ptrianglep
   <\arrowtypea`\arrowtypeb`\arrowtypec;\height>}}
\def\ptrianglep<#1>[#2`#3`#4;#5`#6`#7]{{
\settriparms[#1]
\diagram
\putptrianglep<\arrowtypea`\arrowtypeb`
\arrowtypec;\height>
(0,0)[#2`#3`#4;#5`#6`{#7}]
\enddiagram
}}                                            

\def\putqtrianglep<#1>(#2,#3)[#4`#5`#6;#7`#8`#9]{{%
\settriparms[#1]%
\xpos=#2 \ypos=#3 \advance\ypos by\height
\puthmorphism(\xpos,\ypos)[#4`#5`{#7}]{\height}{\arrowtypea}a%
\putmorphism(\xpos,\ypos)(1,-1)[``{#8}]{\height}{\arrowtypeb}l%
\advance\xpos by\height
\putvmorphism(\xpos,\ypos)[`#6`{#9}]{\height}{\arrowtypec}r%
}}

\def\putqtriangle{\@ifnextchar <{\putqtrianglep}{\putqtrianglep
   <\arrowtypea`\arrowtypeb`\arrowtypec;\height>}}
\def\qtriangle{\@ifnextchar <{\qtrianglep}{\qtrianglep
   <\arrowtypea`\arrowtypeb`\arrowtypec;\height>}}
\def\qtrianglep<#1>[#2`#3`#4;#5`#6`#7]{{
\settriparms[#1]
\width=\height                                
\diagram
\putqtrianglep<\arrowtypea`\arrowtypeb`
\arrowtypec;\height>
(0,0)[#2`#3`#4;#5`#6`{#7}]
\enddiagram
}}

\def\putdtrianglep<#1>(#2,#3)[#4`#5`#6;#7`#8`#9]{{%
\settriparms[#1]%
\xpos=#2 \ypos=#3
\puthmorphism(\xpos,\ypos)[#5`#6`{#9}]{\height}{\arrowtypec}b%
\advance\xpos by \height \advance\ypos by\height
\putmorphism(\xpos,\ypos)(-1,-1)[``{#7}]{\height}{\arrowtypea}l%
\putvmorphism(\xpos,\ypos)[#4``{#8}]{\height}{\arrowtypeb}r%
}}

\def\putdtriangle{\@ifnextchar <{\putdtrianglep}{\putdtrianglep
   <\arrowtypea`\arrowtypeb`\arrowtypec;\height>}}
\def\dtriangle{\@ifnextchar <{\dtrianglep}{\dtrianglep
   <\arrowtypea`\arrowtypeb`\arrowtypec;\height>}}
\def\dtrianglep<#1>[#2`#3`#4;#5`#6`#7]{{
\settriparms[#1]
\width=\height                                
\diagram
\putdtrianglep<\arrowtypea`\arrowtypeb`
\arrowtypec;\height>
(0,0)[#2`#3`#4;#5`#6`{#7}]
\enddiagram
}}

\def\putbtrianglep<#1>(#2,#3)[#4`#5`#6;#7`#8`#9]{{%
\settriparms[#1]%
\xpos=#2 \ypos=#3
\puthmorphism(\xpos,\ypos)[#5`#6`{#9}]{\height}{\arrowtypec}b%
\advance\ypos by\height
\putmorphism(\xpos,\ypos)(1,-1)[``{#8}]{\height}{\arrowtypeb}r%
\putvmorphism(\xpos,\ypos)[#4``{#7}]{\height}{\arrowtypea}l%
}}

\def\putbtriangle{\@ifnextchar <{\putbtrianglep}{\putbtrianglep
   <\arrowtypea`\arrowtypeb`\arrowtypec;\height>}}
\def\btriangle{\@ifnextchar <{\btrianglep}{\btrianglep
   <\arrowtypea`\arrowtypeb`\arrowtypec;\height>}}
\def\btrianglep<#1>[#2`#3`#4;#5`#6`#7]{{
\settriparms[#1]
\width=\height                               
\diagram
\putbtrianglep<\arrowtypea`\arrowtypeb`
\arrowtypec;\height>
(0,0)[#2`#3`#4;#5`#6`{#7}]
\enddiagram
}}

\def\putAtrianglep<#1>(#2,#3)[#4`#5`#6;#7`#8`#9]{{%
\settriparms[#1]%
\xpos=#2 \ypos=#3 {\multiply \height by2
\puthmorphism(\xpos,\ypos)[#5`#6`{#9}]{\height}{\arrowtypec}b}%
\advance\xpos by\height \advance\ypos by\height
\putmorphism(\xpos,\ypos)(-1,-1)[#4``{#7}]{\height}{\arrowtypea}l%
\putmorphism(\xpos,\ypos)(1,-1)[``{#8}]{\height}{\arrowtypeb}r%
}}

\def\putAtriangle{\@ifnextchar <{\putAtrianglep}{\putAtrianglep
   <\arrowtypea`\arrowtypeb`\arrowtypec;\height>}}
\def\Atriangle{\@ifnextchar <{\Atrianglep}{\Atrianglep
   <\arrowtypea`\arrowtypeb`\arrowtypec;\height>}}
\def\Atrianglep<#1>[#2`#3`#4;#5`#6`#7]{{
\settriparms[#1]
\width=\height                                     
\diagram
\putAtrianglep<\arrowtypea`\arrowtypeb`
\arrowtypec;\height>
(0,0)[#2`#3`#4;#5`#6`{#7}]
\enddiagram
}}

\def\putAtrianglepairp<#1>(#2)[#3;#4`#5`#6`#7`#8]{{%
\settripairparms[#1]%
\setpos(#2)%
\settokens`#3`%
\puthmorphism(\xpos,\ypos)[\tokenb`\tokenc`{#7}]{\height}{\arrowtyped}b%
\advance\xpos by\height
\puthmorphism(\xpos,\ypos)[\phantom{\tokenc}`\tokend`{#8}]%
{\height}{\arrowtypee}b%
\advance\ypos by\height
\putmorphism(\xpos,\ypos)(-1,-1)[\tokena``{#4}]{\height}{\arrowtypea}l%
\putvmorphism(\xpos,\ypos)[``{#5}]{\height}{\arrowtypeb}m%
\putmorphism(\xpos,\ypos)(1,-1)[``{#6}]{\height}{\arrowtypec}r%
}}

\def\putAtrianglepair{\@ifnextchar <{\putAtrianglepairp}{\putAtrianglepairp%
   <\arrowtypea`\arrowtypeb`\arrowtypec`\arrowtyped`\arrowtypee;\height>}}
\def\Atrianglepair{\@ifnextchar <{\Atrianglepairp}{\Atrianglepairp%
   <\arrowtypea`\arrowtypeb`\arrowtypec`\arrowtyped`\arrowtypee;\height>}}

\def\Atrianglepairp<#1>[#2;#3`#4`#5`#6`#7]{{
\settripairparms[#1]
\settokens`#2`
\width=\height                                
\diagram
\putAtrianglepairp                            
<\arrowtypea`\arrowtypeb`\arrowtypec`
\arrowtyped`\arrowtypee;\height>
(0,0)[{#2};#3`#4`#5`#6`{#7}]
\enddiagram
}}

\def\putVtrianglep<#1>(#2,#3)[#4`#5`#6;#7`#8`#9]{{%
\settriparms[#1]%
\xpos=#2 \ypos=#3 \advance\ypos by\height {\multiply\height by2
\puthmorphism(\xpos,\ypos)[#4`#5`{#7}]{\height}{\arrowtypea}a}%
\putmorphism(\xpos,\ypos)(1,-1)[`#6`{#8}]{\height}{\arrowtypeb}l%
\advance\xpos by\height \advance\xpos by\height
\putmorphism(\xpos,\ypos)(-1,-1)[``{#9}]{\height}{\arrowtypec}r%
}}

\def\putVtriangle{\@ifnextchar <{\putVtrianglep}{\putVtrianglep
   <\arrowtypea`\arrowtypeb`\arrowtypec;\height>}}
\def\Vtriangle{\@ifnextchar <{\Vtrianglep}{\Vtrianglep
   <\arrowtypea`\arrowtypeb`\arrowtypec;\height>}}
\def\Vtrianglep<#1>[#2`#3`#4;#5`#6`#7]{{
\settriparms[#1]
\width=\height                                 
\diagram
\putVtrianglep<\arrowtypea`\arrowtypeb`
\arrowtypec;\height>
(0,0)[#2`#3`#4;#5`#6`{#7}]
\enddiagram
}}

\def\putVtrianglepairp<#1>(#2)[#3;#4`#5`#6`#7`#8]{{
\settripairparms[#1]%
\setpos(#2)%
\settokens`#3`%
\advance\ypos by\height
\putmorphism(\xpos,\ypos)(1,-1)[`\tokend`{#6}]{\height}{\arrowtypec}l%
\puthmorphism(\xpos,\ypos)[\tokena`\tokenb`{#4}]{\height}{\arrowtypea}a%
\advance\xpos by\height
\puthmorphism(\xpos,\ypos)[\phantom{\tokenb}`\tokenc`{#5}]%
{\height}{\arrowtypeb}a%
\putvmorphism(\xpos,\ypos)[``{#7}]{\height}{\arrowtyped}m%
\advance\xpos by\height
\putmorphism(\xpos,\ypos)(-1,-1)[``{#8}]{\height}{\arrowtypee}r%
}}

\def\putVtrianglepair{\@ifnextchar <{\putVtrianglepairp}{\putVtrianglepairp%
    <\arrowtypea`\arrowtypeb`\arrowtypec`\arrowtyped`\arrowtypee;\height>}}
\def\Vtrianglepair{\@ifnextchar <{\Vtrianglepairp}{\Vtrianglepairp%
    <\arrowtypea`\arrowtypeb`\arrowtypec`\arrowtyped`\arrowtypee;\height>}}
\def\Vtrianglepairp<#1>[#2;#3`#4`#5`#6`#7]{{
\settripairparms[#1]
\settokens`#2`
\diagram
\putVtrianglepairp                             
<\arrowtypea`\arrowtypeb`\arrowtypec`
\arrowtyped`\arrowtypee;\height>
(0,0)[{#2};#3`#4`#5`#6`{#7}]
\enddiagram
}}

\def\putCtrianglep<#1>(#2,#3)[#4`#5`#6;#7`#8`#9]{{%
\settriparms[#1]%
\xpos=#2 \ypos=#3 \advance\ypos by\height
\putmorphism(\xpos,\ypos)(1,-1)[``{#9}]{\height}{\arrowtypec}l%
\advance\xpos by\height \advance\ypos by\height
\putmorphism(\xpos,\ypos)(-1,-1)[#4`#5`{#7}]{\height}{\arrowtypea}l%
{\multiply\height by 2
\putvmorphism(\xpos,\ypos)[`#6`{#8}]{\height}{\arrowtypeb}r}%
}}

\def\putCtriangle{\@ifnextchar <{\putCtrianglep}{\putCtrianglep
    <\arrowtypea`\arrowtypeb`\arrowtypec;\height>}}
\def\Ctriangle{\@ifnextchar <{\Ctrianglep}{\Ctrianglep
    <\arrowtypea`\arrowtypeb`\arrowtypec;\height>}}
\def\Ctrianglep<#1>[#2`#3`#4;#5`#6`#7]{{
\settriparms[#1]
\width=\height                               
\diagram
\putCtrianglep<\arrowtypea`\arrowtypeb`
\arrowtypec;\height>
(0,0)[#2`#3`#4;#5`#6`{#7}]
\enddiagram
}}                                           
\def\putDtrianglep<#1>(#2,#3)[#4`#5`#6;#7`#8`#9]{{%
\settriparms[#1]%
\xpos=#2 \ypos=#3 \advance\xpos by\height \advance\ypos by\height
\putmorphism(\xpos,\ypos)(-1,-1)[``{#9}]{\height}{\arrowtypec}r%
\advance\xpos by-\height \advance\ypos by\height
\putmorphism(\xpos,\ypos)(1,-1)[`#5`{#8}]{\height}{\arrowtypeb}r%
{\multiply\height by 2
\putvmorphism(\xpos,\ypos)[#4`#6`{#7}]{\height}{\arrowtypea}l}%
}}

\def\putDtriangle{\@ifnextchar <{\putDtrianglep}{\putDtrianglep
    <\arrowtypea`\arrowtypeb`\arrowtypec;\height>}}
\def\Dtriangle{\@ifnextchar <{\Dtrianglep}{\Dtrianglep
   <\arrowtypea`\arrowtypeb`\arrowtypec;\height>}}
\def\Dtrianglep<#1>[#2`#3`#4;#5`#6`#7]{{
\settriparms[#1]
\width=\height                              
\diagram
\putDtrianglep<\arrowtypea`\arrowtypeb`
\arrowtypec;\height>
(0,0)[#2`#3`#4;#5`#6`{#7}]
\enddiagram
}}                                          
\def\setrecparms[#1`#2]{\width=#1 \height=#2}%

\def\recursep<#1`#2>[#3;#4`#5`#6`#7`#8]{{\m@th
\width=#1 \height=#2 \settokens`#3`
\settowidth{\tempdimen}{$\tokena$} \ifdim\tempdimen=0pt
  \savebox{\tempboxa}{\hbox{$\tokenb$}}%
  \savebox{\tempboxb}{\hbox{$\tokend$}}%
  \savebox{\tempboxc}{\hbox{$#6$}}%
\else
  \savebox{\tempboxa}{\hbox{$\hbox{$\tokena$}\times\hbox{$\tokenb$}$}}%
  \savebox{\tempboxb}{\hbox{$\hbox{$\tokena$}\times\hbox{$\tokend$}$}}%
  \savebox{\tempboxc}{\hbox{$\hbox{$\tokena$}\times\hbox{$#6$}$}}%
\fi \ypos=\height \divide\ypos by 2 \xpos=\ypos \advance\xpos by
\width \bfig
\putCtrianglep<-1`1`1;\ypos>(0,0)[`\tokenc`;#5`#6`{#7}]%
\puthmorphism(\ypos,0)[\tokend`\usebox{\tempboxb}`{#8}]{\width}{-1}b%
\puthmorphism(\ypos,\height)[\tokenb`\usebox{\tempboxa}`{#4}]{\width}{-1}a%
\advance\ypos by \width
\putvmorphism(\ypos,\height)[``\usebox{\tempboxc}]{\height}1r%
\efig }}

\def\recurse{\@ifnextchar <{\recursep}{\recursep<\width`\height>}}

\def\puttwohmorphisms(#1,#2)[#3`#4;#5`#6]#7#8#9{{%
\puthmorphism(#1,#2)[#3`#4`]{#7}0a \ypos=#2 \advance\ypos by 20
\puthmorphism(#1,\ypos)[\phantom{#3}`\phantom{#4}`#5]{#7}{#8}a
\advance\ypos by -40
\puthmorphism(#1,\ypos)[\phantom{#3}`\phantom{#4}`#6]{#7}{#9}b }}

\def\puttwovmorphisms(#1,#2)[#3`#4;#5`#6]#7#8#9{{%
\putvmorphism(#1,#2)[#3`#4`]{#7}0a \xpos=#1 \advance\xpos by -20
\putvmorphism(\xpos,#2)[\phantom{#3}`\phantom{#4}`#5]{#7}{#8}l
\advance\xpos by 40
\putvmorphism(\xpos,#2)[\phantom{#3}`\phantom{#4}`#6]{#7}{#9}r }}

\def\puthcoequalizer(#1)[#2`#3`#4;#5`#6`#7]#8#9{{%
\setpos(#1)%
\puttwohmorphisms(\xpos,\ypos)[#2`#3;#5`#6]{#8}11%
\advance\xpos by #8
\puthmorphism(\xpos,\ypos)[\phantom{#3}`#4`#7]{#8}1{#9} }}

\def\putvcoequalizer(#1)[#2`#3`#4;#5`#6`#7]#8#9{{%
\setpos(#1)%
\puttwovmorphisms(\xpos,\ypos)[#2`#3;#5`#6]{#8}11%
\advance\ypos by -#8
\putvmorphism(\xpos,\ypos)[\phantom{#3}`#4`#7]{#8}1{#9} }}

\def\putthreehmorphisms(#1)[#2`#3;#4`#5`#6]#7(#8)#9{{%
\setpos(#1) \settypes(#8)
\if a#9 %
     \vertsize{\tempcounta}{#5}%
     \vertsize{\tempcountb}{#6}%
     \ifnum \tempcounta<\tempcountb \tempcounta=\tempcountb \fi
\else
     \vertsize{\tempcounta}{#4}%
     \vertsize{\tempcountb}{#5}%
     \ifnum \tempcounta<\tempcountb \tempcounta=\tempcountb \fi
\fi \advance \tempcounta by 60
\puthmorphism(\xpos,\ypos)[#2`#3`#5]{#7}{\arrowtypeb}{#9}
\advance\ypos by \tempcounta
\puthmorphism(\xpos,\ypos)[\phantom{#2}`\phantom{#3}`#4]{#7}{\arrowtypea}{#9}
\advance\ypos by -\tempcounta \advance\ypos by -\tempcounta
\puthmorphism(\xpos,\ypos)[\phantom{#2}`\phantom{#3}`#6]{#7}{\arrowtypec}{#9}
}}

\def\setarrowtoks[#1`#2`#3`#4`#5`#6]{%
\def\toka{#1}
\def\tokb{#2}
\def\tokc{#3}
\def\tokd{#4}
\def\toke{#5}
\def\tokf{#6}
}
\def\hex{\@ifnextchar <{\hexp}{\hexp<1000`400>}}
\def\hexp<#1`#2>[#3`#4`#5`#6`#7`#8;#9]{%
\setarrowtoks[#9] \yext=#2 \advance \yext by #2 \xext=#1
\advance\xext by \yext \bfig
\putCtriangle<-1`0`1;#2>(0,0)[`#5`;\tokb``\tokd] \xext=#1
\yext=#2 \advance \yext by #2
\putsquare<1`0`0`1;\xext`\yext>(#2,0)[#3`#4`#7`#8;\toka```\tokf]
\advance \xext by #2
\putDtriangle<0`1`-1;#2>(\xext,0)[`#6`;`\tokc`\toke] \efig }

\makeatother

\input{tcilatex}
\begin{document}

\title{Geometrical Bioelectrodynamics}
\author{Vladimir G. Ivancevic and Tijana T. Ivancevic}
\date{}
\maketitle

\begin{abstract}

This paper proposes rigorous geometrical treatment of
\emph{bioelectrodynamics}, underpinning two fast--growing
biomedical research fields: \emph{bioelectromagnetism}, which
deals with the ability of life to produce its own
electromagnetism, and \emph{bioelectromagnetics}, which deals with
the effect on life from external electromagnetism.
\newline

\noindent\textbf{Keywords:} Bioelectrodynamics, exterior
geometrical machinery, Dirac--Feynman quantum electrodynamics,
functional electrical stimulation
\end{abstract}

\tableofcontents

\section{Introduction}

\emph{Bioelectromagnetism}, sometimes equated with
\emph{bioelectricity}, bioelectromagnetism refers to the
electrical, magnetic or electromagnetic fields produced by living
cells, tissues or organisms. Examples include the cell potential
of cell membranes and the electric currents that flow in nerves
and muscles, as a result of action potentials. It is studied
primarily through the techniques of \emph{electrophysiology} (see,
e.g. \cite{Brazier}), which is the study of the electrical
properties of biological cells and tissues. It involves
measurements of voltage change or electrical current flow on a
wide variety of scales from single ion channel proteins to whole
tissues like the heart. In neuroscience, it includes measurements
of the electrical activity of neurons, and particularly
\emph{neural action potentials}, or neural spikes, pioneered by
five celebrated Hodgkin-Huxley papers in 1952
\cite{HH1,HH2,HH3,HH4,HH5} (formalized in \cite{HH5} by a
1963-Nobel-Prize winning mathematical model), which are pulse-like
waves of voltage that travel along several types of cell
membranes. The best-understood example is generated on the
membrane of the axon of a neuron, but also appears in other types
of excitable cells, such as cardiac muscle cells, and even plant
cells.\footnote{The resting voltage across the axonal membrane is
typically -70 millivolts (mV), with the inside being more negative
than the outside. As an action potential passes through a point,
this voltage rises to roughly +40 mV in one millisecond, then
returns to -70 mV. The action potential moves rapidly down the
axon, with a conduction velocity as high as 100 meters/second (224
miles per hour). Because they are able to transmit information so
fast, the flow of action potentials is a very efficient form of
data transmission, considering that each neuron the signal passes
through can be up to a meter in length. An action potential is
provoked on a patch of membrane when the membrane is depolarized
sufficiently, i.e., when the voltage of the cell's interior
relative to the cell's exterior is raised above a threshold. Such
a depolarization opens voltage-sensitive channels, which allows
current to flow into the axon, further depolarizing the membrane.
This will cause the membrane to `fire', initiating a positive
feedback loop that suddenly and rapidly causes the voltage inside
the axon to become more positive. After this rapid rise, the
membrane voltage is restored to its resting value by a combination
of effects: the channels responsible for the initial inward
current are inactivated, while the raised voltage opens other
voltage-sensitive channels that allow a compensating outward
current. Because of the positive feedback, an action potential is
all-or-none; there are no partial action potentials. In neurons, a
typical action potential lasts for just a few thousandths of a
second at any given point along their length. The passage of an
action potential can leave the ion channels in a non-equilibrium
state, making them more difficult to open, and thus inhibiting
another action potential at the same spot: such an axon is said to
be refractory.} The principal ions involved in an action potential
are sodium and potassium cations; sodium ions enter the cell, and
potassium ions leave, restoring equilibrium. Relatively few ions
need to cross the membrane for the membrane voltage to change
drastically. The ions exchanged during an action potential,
therefore, make a negligible change in the interior and exterior
ionic concentrations. The few ions that do cross are pumped out
again by the continual action of the sodium–potassium pump, which,
with other ion transporters, maintains the normal ratio of ion
concentrations across the membrane.\footnote{The action potential
`travels' along the axon without fading out because the signal is
regenerated at each patch of membrane. This happens because an
action potential at one patch raises the voltage at nearby
patches, depolarizing them and provoking a new action potential
there. In unmyelinated neurons, the patches are adjacent, but in
myelinated neurons, the action potential `hops' between distant
patches, making the process both faster and more efficient. The
axons of neurons generally branch, and an action potential often
travels along both forks from a branch point. The action potential
stops at the end of these branches, but usually causes the
secretion of neurotransmitters at the synapses that are found
there. These neurotransmitters bind to receptors on adjacent
cells. These receptors are themselves ion channels, although—in
contrast to the axonal channels—they are generally opened by the
presence of a neurotransmitter, rather than by changes in voltage.
The opening of these receptor channels can help to depolarize the
membrane of the new cell (an excitatory channel) or work against
its depolarization (an inhibitory channel). If these
depolarizations are sufficiently strong, they can provoke another
action potential in the new cell. \par The flow of currents within
an axon can be described quantitatively by \emph{cable theory}
\cite{Rall} (as demonstrated by Hodgkin \cite{Hodgkin46}). In
simple cable theory, the neuron is treated as an electrically
passive, perfectly cylindrical transmission cable, which can be
described by a PDE: $$\tau \frac{\partial V}{\partial t} =
\lambda^{2} \frac{\partial^{2} V}{\partial x^{2}} - V,$$ where
$V(x,t)$ is the voltage across the membrane at a time $t$ and a
position $x$ along the length of the neuron, and where $\lambda$
and $\tau$ are the characteristic length and time scales on which
those voltages decay in response to a stimulus.} Calcium cations
and chloride anions are involved in a few types of action
potentials, such as the \emph{cardiac action
potential}\footnote{The cardiac action potential differs from the
neuronal action potential by having an extended plateau, in which
the membrane is held at a high voltage for a few hundred
milliseconds prior to being repolarized by the potassium current
as usual. This plateau is due to the action of slower calcium
channels opening and holding the membrane voltage near their
equilibrium potential even after the sodium channels have
inactivated. The cardiac action potential plays an important role
in coordinating the contraction of the heart. The cardiac cells of
the sinoatrial node provide the pacemaker potential that
synchronizes the heart. The action potentials of those cells
propagate to and through the atrioventricular node (AV node),
which is normally the only conduction pathway between the atria
and the ventricles. Action potentials from the AV node travel
through the bundle of His and thence to the Purkinje fibers.
Conversely, anomalies in the cardiac action potential, whether due
to a congenital mutation or injury, can lead to human pathologies,
especially arrhythmias (see, e.g. \cite{Kleber}).} \cite{Noble}.

The action potential in a normal skeletal muscle cell is similar
to the action potential in neurons \cite{NatBio,GaneshSprSml}.
Action potentials result from the depolarization of the cell
membrane (the sarcolemma), which opens voltage-sensitive sodium
channels; these becomes inactivated and the membrane is
repolarized through the outward current of potassium ions. The
resting potential prior to the action potential is typically
-90mV, somewhat more negative than typical neurons. The muscle
action potential lasts roughly 2–4 ms, the absolute refractory
period is roughly 1–3 ms, and the conduction velocity along the
muscle is roughly 5 m/s. The action potential releases calcium
ions that free up the tropomyosin and allow the muscle to
contract. Muscle action potentials are provoked by the arrival of
a pre-synaptic neuronal action potential at the neuromuscular
junction. There output is muscle force generation.

Bioelectromagnetic activity of the brain is measured by
non-invasive \emph{electroencephalography} (EEG), usually recorded
from electrodes placed on the scalp\footnote{The data measured by
the scalp EEG are used for clinical and research purposes. A
technique similar to the EEG is intracranial EEG (icEEG), also
referred to as subdural EEG (sdEEG) and electrocorticography
(ECoG). These terms refer to the recording of activity from the
surface of the brain (rather than the scalp). Because of the
filtering characteristics of the skull and scalp, icEEG activity
has a much higher spatial resolution than surface EEG.} (see, e.g.
\cite{Epstein}). Normal brain wave patterns are classified by the
frequency range as: delta ($<4$Hz),\footnote{Delta is the highest
in amplitude and the slowest in frequency. It is seen normally in
adults in slow wave sleep. It is also seen normally in babies. It
may occur focally with subcortical lesions and in general
distribution with diffuse lesions, metabolic encephalopathy
hydrocephalus or deep midline lesions. It is usually most
prominent frontally in adults (e.g. Frontal Intermittent Rhythmic
Delta) and posteriorly in children e.g. Occipital Intermittent
Rhythmic Delta). } theta ($4-7$Hz),\footnote{Theta is seen
normally in young children. It may be seen in drowsiness or
arousal in older children and adults, as well as in meditation.}
alpha ($7-12$Hz),\footnote{Alpha is seen in the posterior regions
of the head on both sides, being higher in amplitude on the
dominant side. It is brought out by closing the eyes and by
relaxation. It was noted to attenuate with eye opening or mental
exertion. This activity is now referred to as `posterior basic
rhythm', the `posterior dominant rhythm' or the `posterior alpha
rhythm', and is actually slower than 8Hz in young children
(therefore technically in the theta range). In addition, there are
two other normal alpha rhythms that are typically discussed: the
Mu rhythm (which is an alpha--range activity that is seen over the
sensorimotor cortex and characteristically attenuates with
movement of the contralateral arm) and a temporal `third rhythm'.}
beta ($12-30$Hz),\footnote{Beta is usually seen on both sides of
the scalp in symmetrical distribution and is most evident
frontally. Low amplitude beta with multiple and varying
frequencies is often associated with active, busy or anxious
thinking and active concentration.} and gamma
($30>$Hz).\footnote{Because of the filtering properties of the
skull and scalp, gamma rhythms can only be recorded from
electrocorticography or possibly with magnetoencephalography.
Gamma rhythms are thought to represent binding of different
populations of neurons together into a network for the purpose of
carrying out a certain cognitive or motor function.}

Bioelectromagnetic activity of the heart is measured by
non-invasive \emph{electrocardiography} (ECG or EKG), usually
recorded from electrodes placed on the skin of a thorax
 (see, e.g. \cite{Braunwald}).\footnote{Electrodes on different sides of the heart measure the
activity of different parts of the heart muscle. An ECG displays
the voltage between pairs of these electrodes, and the muscle
activity that they measure, from different directions, also
understood as vectors. This display indicates the overall rhythm
of the heart, and weaknesses in different parts of the heart
muscle. It is the best way to measure and diagnose abnormal
rhythms of the heart, particularly abnormal rhythms caused by
damage to the conductive tissue that carries electrical signals,
or abnormal rhythms caused by levels of dissolved salts
(electrolytes), such as potassium, that are too high or low. In
myocardial infarction (MI), the ECG can identify damaged heart
muscle. But it can only identify damage to muscle in certain
areas, so it can't rule out damage in other areas. The ECG cannot
reliably measure the pumping ability of the heart; for which
ultrasound-based (echocardiography) or nuclear medicine tests are
used.} Human heartbeats originate from the sinoatrial node (SA
node) near the right atrium of the heart. Modified muscle cells
contract, sending a signal to other muscle cells in the heart to
contract. The signal spreads to the atrioventricular node (AV
node). Signals carried from the AV node, slightly delayed, through
bundle of His fibers and Purkinje fibers cause the ventricles to
contract simultaneously. ECG measures changes in electrical
potential across the heart, and can detect the contraction pulses
that pass over the surface of the heart. There are three slow,
negative changes, known as P, R, and T. Positive deflections are
the Q and S waves. The P wave represents the contraction impulse
of the atria, the T wave the ventricular contraction.\footnote{The
baseline voltage of the electrocardiogram is known as the
\textit{isoelectric line}. A typical ECG tracing of a normal
heartbeat (or, cardiac cycle) consists of a P wave, a QRS complex
and a T wave (see, e.g. \cite{Mark}). A small U wave is normally
visible in 50 to 75\% of ECGs. During normal atrial
depolarization, the mean electrical vector is directed from the SA
node towards the AV node, and spreads from the right atrium to the
left atrium. This turns into the P wave on the EKG, which is
upright in II, III, and aVF (since the general electrical activity
is going toward the positive electrode in those leads), and
inverted in aVR (since it is going away from the positive
electrode for that lead). The relationship between P waves and QRS
complexes helps distinguish various cardiac arrhythmias. The shape
and duration of the P waves may indicate atrial enlargement. The
PR interval is measured from the beginning of the P wave to the
beginning of the QRS complex. It is usually 0.12 to 0.20 s (120 to
200 ms). A prolonged PR interval may indicate a first degree heart
block. A short PR interval may indicate an accessory pathway that
leads to early activation of the ventricles, such as seen in
Wolff--Parkinson--White syndrome. A variable PR interval may
indicate other types of heart block. PR segment depression may
indicate atrial injury or pericarditis. The QRS complex is a
structure on the ECG that corresponds to the depolarization of the
ventricles. Because the ventricles contain more muscle mass than
the atria, the QRS complex is larger than the P wave. In addition,
because the His/Purkinje system coordinates the depolarization of
the ventricles, the QRS complex tends to look `spiked' rather than
rounded due to the increase in conduction velocity. A normal QRS
complex is 0.06 to 0.10 sec (60 to 100 ms) in duration. Not every
QRS complex contains a Q wave, an R wave, and an S wave. By
convention, any combination of these waves can be referred to as a
QRS complex. The duration, amplitude, and morphology of the QRS
complex is useful in diagnosing cardiac arrhythmias, conduction
abnormalities, ventricular hypertrophy, myocardial infarction,
electrolyte derangements, and other disease states. Q waves can be
normal (physiological) or pathological. The ST segment connects
the QRS complex and the T wave and has a duration of 0.08 to 0.12
s (80 to 120 ms). It starts at the J point (junction between the
QRS complex and ST segment) and ends at the beginning of the T
wave. However, since it is usually difficult to determine exactly
where the ST segment ends and the T wave begins, the relationship
between the ST segment and T wave should be examined together. The
typical ST segment duration is usually around 0.08 s (80 ms). It
should be essentially level with the PR and TP segment. The normal
ST segment has a slight upward concavity. Flat, down--sloping, or
depressed ST segments may indicate coronary ischemia. ST segment
elevation may indicate myocardial infarction.}

Bioelectromagnetic activity of skeletal muscles is measured by
non-invasive \emph{electromyography} (EMG), usually recorded from
surface electrodes placed on the skin above working
muscles.\footnote{In contrast, to perform invasive intramuscular
EMG, a needle electrode is inserted through the skin into the
muscle tissue. The insertional activity provides valuable
information about the state of the muscle and its innervating
nerve.} An electromyograph detects the electrical potential
generated by muscle cells, both when they contract or are at
rest.\footnote{The electrical source is the muscle membrane
potential of about -70mV. Measured EMG potentials range between
less than 50 µV and up to 20 to 30 mV, depending on the muscle
under observation. Typical repetition rate of muscle unit firing
is about 7–20 Hz, depending on the size of the muscle (eye muscles
versus seat (gluteal) muscles), previous axonal damage and other
factors. Damage to motor units can be expected at ranges between
450 and 780 mV.} A surface electrode may be used to monitor the
general picture of muscle activation (see, e.g. \cite{Cram}), as
opposed to the activity of only a few fibres as observed using a
needle. This technique is used in a number of settings; for
example, in the physiotherapy clinic, muscle activation is
monitored using surface EMG and patients have an auditory or
visual stimulus to help them know when they are activating the
muscle (biofeedback).\footnote{A motor unit is defined as one
motor neuron and all of the muscle fibers it innervates. When a
motor unit fires, the impulse (called an action potential) is
carried down the motor neuron to the muscle. The area where the
nerve contacts the muscle is called the neuromuscular junction, or
the motor end plate. After the action potential is transmitted
across the neuromuscular junction, an action potential is elicited
in all of the innervated muscle fibers of that particular motor
unit. The sum of all this electrical activity is known as a motor
unit action potential (MUAP). This electrophysiologic activity
from multiple motor units is the signal typically evaluated during
an EMG. The composition of the motor unit, the number of muscle
fibres per motor unit, the metabolic type of muscle fibres and
many other factors affect the shape of the motor unit potentials
in the myogram. Nerve conduction testing is also often done at the
same time as an EMG in order to diagnose neurological diseases.}
EMG signals are essentially made up of superimposed motor unit
action potentials (MUAPs) from several motor units. For a thorough
analysis, the measured EMG signals can be decomposed into their
constituent MUAPs. MUAPs from different motor units tend to have
different characteristic shapes, while MUAPs recorded by the same
electrode from the same motor unit are typically
similar.\footnote{Notably MUAP size and shape depend on where the
electrode is located with respect to the fibers and so can appear
to be different if the electrode moves position. EMG decomposition
is non-trivial, although many methods have been proposed (see,
e.g. \cite{Reaz}).}

Bioelectromagnetism should not to be confused with the similar
word \emph{bioelectromagnetics}, which deals with the effect on
life from external electromagnetism (see, e.g. \cite{Carpenter}).
Supported by the Bioelectromagnetics Society (with Dr. Stefan
Engstr\"om as President) and the scientific journal with the same
name, bioelectromagnetics is the study of how electromagnetic
fields interact with and influence biological processes. Common
areas of investigation include the mechanism of animal migration
and navigation using the geomagnetic field, studying the potential
effects of man-made sources of electromagnetic fields, such as
those produced by the power distribution system and mobile phones,
and developing novel therapies to treat various conditions.

Most importantly, in recent years a growing number of people have
begun to use mobile phone technology. This phenomenon has raised
questions and doubts about possible effects on users' brains. The
recent literature review presented in \cite{Valentini} has focused
on the human electrophysiological and neuro-metabolic effects of
mobile phone (MP)- related electromagnetic fields (EMFs) published
in the last 10 years. All relevant papers have been reported and,
subsequently, a literature selection has been carried out by
taking several criteria into account, such as: blind techniques,
randomization or counter-balancing of conditions and subjects,
detail of exposure characteristics and the statistical analyses
used. As a result, only the studies meeting the selection criteria
have been described, evaluated and discussed further. This review
has provided a clear scenario of the most reliable experiments
carried out over the last decade and to offer a critical point of
view in their evaluation. Its conclusion is that MP-EMFs may
influence normal physiology through changes in cortical
excitability and that in future research particular care should be
dedicated to both methodological and statistical control, the most
relevant criteria in this research field.

More generally, epidemiological studies of radio frequency (RF)
exposures and human cancers include studies of military and
civilian occupational groups, people who live near television and
radio transmitters, and users of mobile phones. Many types of
cancer have been assessed in \cite{Elwood}, with particular
attention given to leukemia and brain tumors. The presented
epidemiological results fall short of the strength and consistency
of evidence that is required to come to a conclusion that RF
emissions are a cause of human cancer. Although the
epidemiological evidence in total suggests no increased risk of
cancer, the results cannot be unequivocally interpreted in terms
of cause and effect. The results are inconsistent, and most
studies are limited by lack of detail on actual exposures, short
follow-up periods, and the limited ability to deal with other
relevant factors. For these reasons, the studies are unable to
confidently exclude any possibility of an increased risk of
cancer. Further research to clarify the situation is justified.
Priorities include further studies of leukemia in both adults and
children, and of cranial tumors in relationship to mobile phone
use.

In a recent technical paper \cite{Vincze}, a new theoretical
treatment of ion resonance phenomena was formulated around the
\emph{Lorentz force equation}. This equation defines the motion of
a particle of mass $m$ and charge $q$ moving with a mean time
$\tau$ in a joint electric field $\mathbf{E}$ and magnetic field
$\mathbf{B}$
\begin{equation}
m\mathbf{\dot{v}}+m\mathbf{v}/\tau =q(\mathbf{E}+\mathbf{v}\times
\mathbf{B}), \label{LorFor}
\end{equation}
where overdot means time derivative and $\times$ is the vector
cross--product. The authors derive an expression for the velocity
of a damped ion with arbitrary $q/m$ under the influence of the
Lorentz force on the right--hand side of (\ref{LorFor}). Series
solutions to the differential equations reveal transient responses
as well as resonance-like terms. An important result in this field
is that the expressions for ionic drift velocity include a
somewhat similar Bessel function dependence as was previously
obtained for the transition probability in parametric resonance.
The authors have found an explicit effect due to damping, so that
previous Bessel dependence now occurs as a subset of a more
general solution, including not only the magnetic field AC/DC
ratio as an independent variable, but also the ratio of the
cyclotronic frequency $\Omega$ to the applied AC frequency
$\omega$. The authors hypothesize that the selectively enhanced
drift velocity predicted in their model can explain ICR-like
phenomena as resulting from increased interaction probabilities in
the vicinity of ion channel gates.

This paper proposes rigorous geometrical
\emph{bioelectrodynamics}, which we define to be the common field
theory underpinning both bioelectromagnetism and
bioelectromagnetics.

\section{Exterior Geometrical Machinery in 4D}

\subsection{From Green to Stokes}

Recall that \textit{Green's Theorem} in the region $C\ $in $(x,y)-$plane $%
\mathbb{R}^{2}$\ connects a line integral $\oint_{\partial C}$ (over the
boundary $\partial C$ of $C)$ with a double integral $\iint_{C}$ over $C$
(see e.g., \cite{MarsVec})
\begin{equation*}
\oint_{\partial C}Pdx+Qdy=\iint_{C}\left( \frac{\partial Q}{\partial x}-%
\frac{\partial P}{\partial y}\right) dxdy.
\end{equation*}%
In other words, if we define two differential forms (integrands of $%
\oint_{\partial C}$ and $\iint_{C}$) as
\begin{eqnarray*}
\text{1--form} &:&\mathbf{A}=Pdx+Qdy,\qquad \text{and} \\
\text{2--form} &:&\mathbf{dA}=\left( \frac{\partial Q}{\partial x}-\frac{%
\partial P}{\partial y}\right) dxdy,
\end{eqnarray*}%
(where $\mathbf{d}$ denotes the \textit{exterior derivative} that makes a $%
(p+1)-$form out of a $p-$form, see next subsection), then we can rewrite
Green's theorem as \textit{Stokes' theorem:}
\begin{equation*}
\int_{\partial C}\mathbf{A}=\int_{C}\mathbf{dA}.
\end{equation*}%
The integration domain $C$ is in topology called a \textit{chain}, and $%
\partial C$ is a 1D boundary of a 2D chain $C$. In general, \emph{the
boundary of a boundary is zero} (see \cite{MTW,CW}), that is, $\partial
(\partial C)=0$, or formally $\partial ^{2}=0$.

\subsection{Exterior derivative}

\begin{figure}[tbp]
\centerline{\includegraphics[height=13cm]{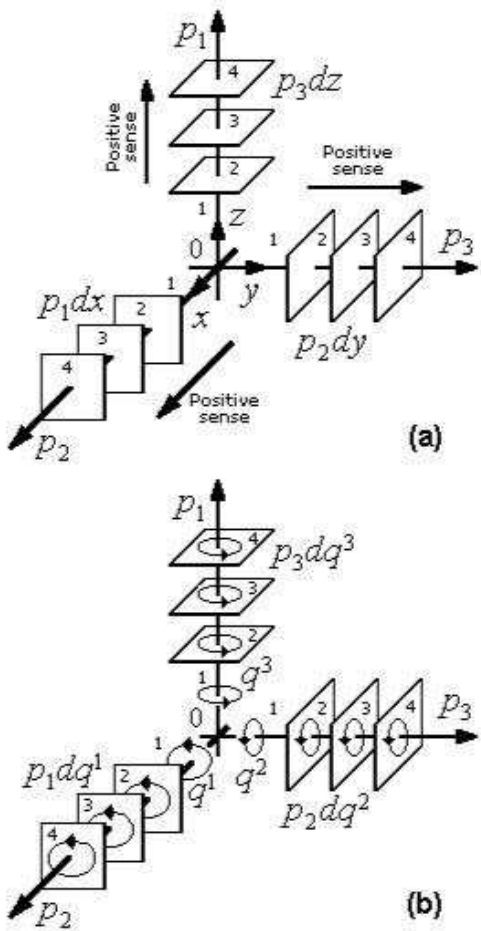}}
\caption{Basis vectors and one-forms in Euclidean $\mathbb{R}^{3}-$space:
(a) Translational case; and (b) Rotational case \protect\cite{VladSIAM}. For
the same geometry in $\mathbb{R}^{3}$, see \protect\cite{MTW}.}
\label{Basis}
\end{figure}

\begin{figure}[tbp]
\centerline{\includegraphics[height=13cm]{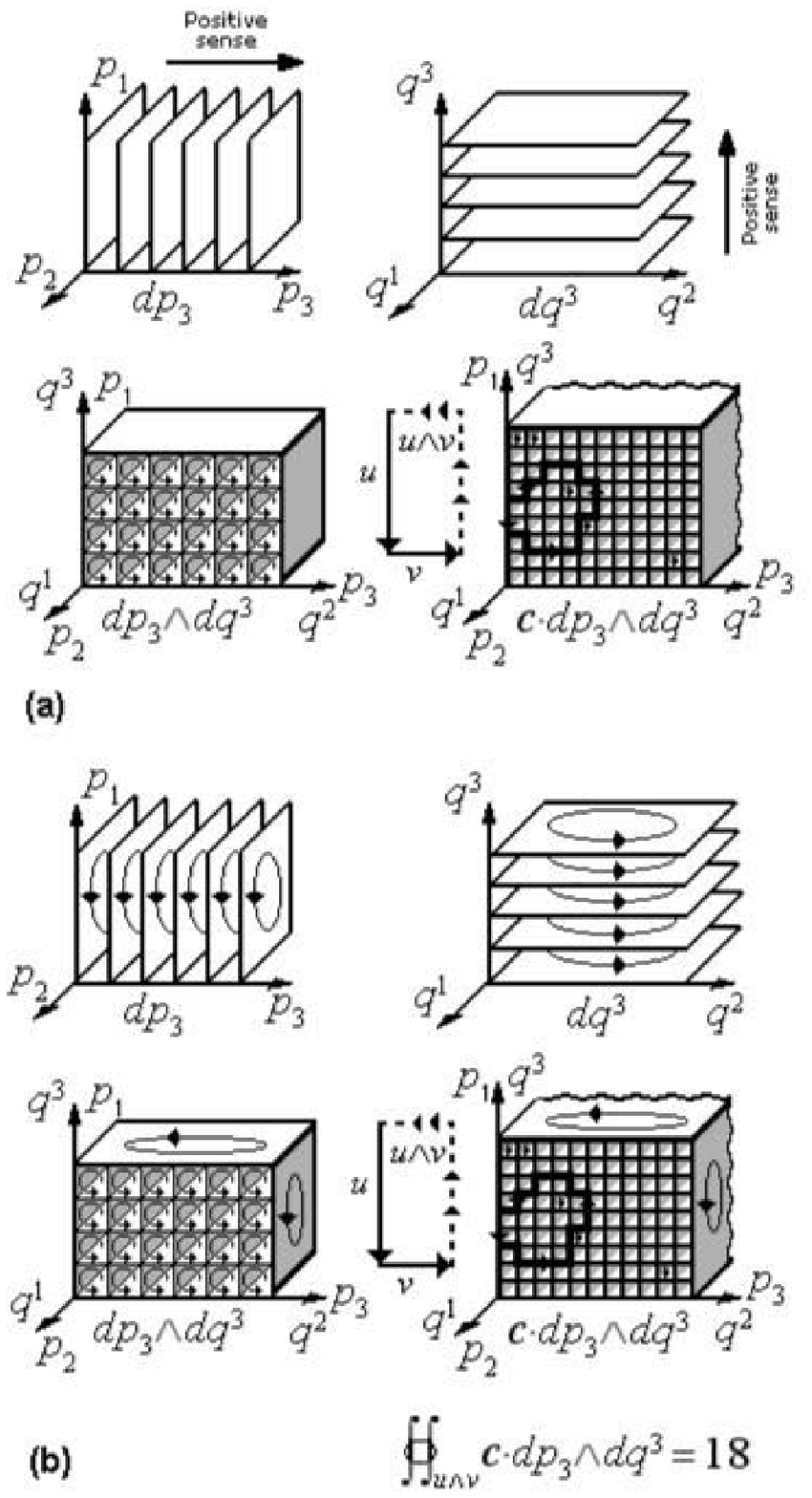}}
\caption{Fundamental two--form and its flux in $\mathbb{R}^3$: (a)
Translational case; (b) Rotational case. In both cases the flux through the
plane $u\wedge v$ is defined as $\protect\int\protect\int_{u\wedge v}
c\,dp_idq^i$ and measured by the number of tubes crossed by the circulation
oriented by $u\wedge v$ \protect\cite{VladSIAM}. For the same geometry in $%
\mathbb{R}^3$, see \protect\cite{MTW}.}
\label{2form}
\end{figure}

The exterior derivative $\mathbf{d}$ is a generalization of ordinary vector
differential operators (\textsl{grad}, \textsl{div} and \textsl{curl} see
e.g., \cite{Flanders}) that transforms $p-$forms $\mathbf{\omega }$ into $%
(p+1)-$forms $\mathbf{d\omega }$ (see next subsection), with the main
property: $\mathbf{dd}=\mathbf{d}^{2}=0$, so that in $\mathbb{R}^{3}$ we
have (see Figures \ref{Basis} and \ref{2form})

\begin{itemize}
\item any scalar function $f=f(x,y,z)$ is a 0--form;

\item the gradient $\mathbf{d}f=\mathbf{\omega }$ of any smooth function $f$
is a 1--form
\begin{equation*}
\mathbf{\omega =d}f=\frac{\partial f}{\partial x}dx+\frac{\partial f}{%
\partial y}dy+\frac{\partial f}{\partial z}dz;
\end{equation*}

\item the curl $\mathbf{\alpha =d\omega}$ of any smooth 1--form $\mathbf{%
\omega }$ is a 2--form
\begin{eqnarray*}
&&\mathbf{\alpha =d\omega} =\left( \frac{\partial R}{\partial y}-\frac{%
\partial Q}{\partial z}\right) dydz+\left( \frac{\partial P}{\partial z}-%
\frac{\partial R}{\partial x}\right) dzdx+\left( \frac{\partial Q}{\partial x%
}-\frac{\partial P}{\partial y}\right) dxdy; \\
&&\text{if }\mathbf{\omega =d}f\Rightarrow \mathbf{\alpha =dd}f=0.
\end{eqnarray*}

\item the divergence $\mathbf{\beta =d\alpha}$ of any smooth 2--form $%
\mathbf{\alpha } $ is a 3--form
\begin{equation*}
\mathbf{\beta =d\alpha} =\left( \frac{\partial A}{\partial x}+\frac{\partial
B}{\partial y}+\frac{\partial C}{\partial z}\right) dxdydz;\qquad\text{if }%
\mathbf{\alpha =d\omega}\Rightarrow \mathbf{\beta =dd\omega}=0.
\end{equation*}
\end{itemize}

In general, for any two smooth functions $f=f(x,y,z)$ and $g=g(x,y,z)$, the
exterior derivative $\mathbf{d}$\ obeys the \emph{Leibniz rule} \cite%
{GaneshSprBig,GaneshADG}:
\begin{equation*}
\mathbf{d}(fg)=g\,\mathbf{d}f+f\,\mathbf{d}g,
\end{equation*}%
and the \textit{chain rule}:
\begin{equation*}
\mathbf{d}\left( g(f)\right) =g^{\prime }(f)\,\mathbf{d}f.
\end{equation*}

\subsection{Exterior $p-$forms in $\mathbb{R}^{4}$}

In general, given a so--called 4D \textit{coframe}, that is a set of
coordinate differentials $\{dx^{i}\}\in \mathbb{R}^{4}$, we can define the
space of all $p-$forms, denoted $\Omega ^{p}(\mathbb{R}^{4})$, using the
exterior derivative $\mathbf{d}:\Omega ^{p}(\mathbb{R}^{4})\rightarrow
\Omega ^{p+1}(\mathbb{R}^{4})$ and Einstein's summation convention over
repeated indices (e.g., $A_{i}\,dx^{i}=\sum_{i=0}^{3}A_{i}\,dx^{i}$), we
have:

\begin{description}
\item[1--form] -- a generalization of the Green's 1--form $Pdx+Qdy$,
\begin{equation*}
\mathbf{A}=A_{i}\,dx^{i}\in \Omega ^{1}(\mathbb{R}^{4}).
\end{equation*}%
For example, in 4D electrodynamics, $\mathbf{A}$ represents electromagnetic
(co)vector potential.

\item[2--form] -- generalizing the Green's 2--form $(\partial _{x}Q-\partial
_{y}P)\,dxdy$ ~$($with $\partial _{j}=\partial /\partial x^{j}$),
\begin{eqnarray*}
\mathbf{B} &=&\mathbf{dA}\in \Omega ^{2}(\mathbb{R}^{4}),\qquad \text{with
components} \\
\mathbf{B} &=&\frac{1}{2}B_{ij}\,dx^{i}\wedge dx^{j},\qquad \text{or} \\
\mathbf{B} &=&\partial _{j}A_{i}\,dx^{j}\wedge dx^{i},\qquad \text{so that}
\\
B_{ij} &=&-2\partial _{j}A_{i}=\partial _{i}A_{j}-\partial _{j}A_{i}=-B_{ji}.
\end{eqnarray*}%
where $\wedge $ is the anticommutative exterior (or, `wedge') product of two
differential forms; given a $p-$form $\mathbf{\alpha }\in \Omega ^{p}(%
\mathbb{R}^{4})$ and a $q-$form $\mathbf{\beta }\in \Omega ^{q}(\mathbb{R}%
^{4}),$ their exterior product is a $(p+q)-$form $\mathbf{\alpha }\wedge
\mathbf{\beta }\in \Omega ^{p+q}(\mathbb{R}^{4})$; e.g., if we have two
1--forms $\mathbf{a}=a_{i}dx^{i}$, and $\mathbf{b}=b_{i}dx^{i}$, their wedge
product $\mathbf{a}\wedge \mathbf{b}$ is a 2--form $\mathbf{\alpha }$ given
by
\begin{equation*}
\mathbf{\alpha }=\mathbf{a}\wedge \mathbf{b}=a_{i}b_{j}\,dx^{i}\wedge
dx^{j}=-a_{i}b_{j}\,dx^{j}\wedge dx^{i}=-\mathbf{b}\wedge \mathbf{a}.
\end{equation*}%
The exterior product $\wedge $ is related to the exterior derivative $%
\mathbf{d}$, by
\begin{equation*}
\mathbf{d}(\mathbf{\alpha }\wedge \mathbf{\beta })=\mathbf{d\alpha }\wedge
\mathbf{\beta }+(-1)^{p}\mathbf{\alpha }\wedge \mathbf{d\beta }.\text{ \ }
\end{equation*}

\item[3--form]
\begin{eqnarray*}
\mathbf{C} &=&\mathbf{dB~(}=\mathbf{ddA}\equiv 0)\in \Omega ^{3}(\mathbb{R}%
^{4}),\qquad \text{with components} \\
\mathbf{C} &=&\frac{1}{3!}C_{ijk}\,dx^{i}\wedge dx^{j}\wedge dx^{k},\qquad
\text{or} \\
\mathbf{C} &=&\partial _{k}B_{[ij]}\,dx^{k}\wedge dx^{i}\wedge dx^{j},\qquad
\text{so that} \\
C_{ijk} &=&-6\partial _{k}B_{[ij]},\qquad \text{where }B_{[ij]}\text{ is the
skew--symmetric part of }B_{ij}.
\end{eqnarray*}%
For example, in the 4D electrodynamics, $\mathbf{B}$ represents
the field 2--form \emph{Faraday}, or the Li\'{e}nard--Wiechert
2--form (in the next section we will use the standard symbol
$\mathbf{F}$ instead of $\mathbf{B}$) satisfying the sourceless
magnetic Maxwell's equation,
\begin{equation*}
\text{Bianchi identity}:\quad \mathbf{dB}=0,\quad \text{in components\quad }%
\partial _{k}B_{[ij]}=0.
\end{equation*}

\item[4--form]
\begin{eqnarray*}
\mathbf{D} &=&\mathbf{dC~(}=\mathbf{ddB}\equiv 0)\in \Omega ^{4}(\mathbb{R}%
^{4}),\qquad \text{with components} \\
\mathbf{D} &=&\partial _{l}C_{[ijk]}\,dx^{l}\wedge dx^{i}\wedge dx^{j}\wedge
dx^{k},\qquad \text{or} \\
\mathbf{D} &=&\frac{1}{4!}D_{ijkl}\,dx^{i}\wedge dx^{j}\wedge dx^{k}\wedge
dx^{l},\qquad \text{so that} \\
D_{ijkl} &=&-24\partial _{l}C_{[ijk]}.
\end{eqnarray*}
\end{description}

\subsection{Stokes Theorem in $\mathbb{R}^{4}$}

Generalization of the Green's theorem in the plane (and all other integral
theorems from vector calculus) is the Stokes Theorem for the $p-$form $%
\mathbf{\omega }$, in the $n$D domain $C$ (which is a $p-$chain with a $%
(p-1)-$boundary $\partial C$, see next subsection)
\begin{equation*}
\int_{\partial C}\mathbf{\omega }=\int_{C}\mathbf{d\omega }.
\end{equation*}%
In the 4D Euclidean space $\mathbb{R}^{4}$ we have the following three
particular cases of the Stokes theorem, related to the subspaces $C$ of $%
\mathbb{R}^{4}$:

The 2D Stokes theorem:

\begin{equation*}
\int_{\partial C^{2}}\mathbf{A}=\int_{C^{2}}\mathbf{B}.
\end{equation*}

The 3D Stokes theorem:

\begin{equation*}
\int_{\partial C^{3}}\mathbf{B}=\int_{C^{3}}\mathbf{C}.
\end{equation*}

The 4D Stokes theorem:

\begin{equation*}
\int_{\partial C^{4}}\mathbf{C}=\int_{C^{4}}\mathbf{D}.
\end{equation*}

\subsection{Exact and closed $p-$forms and $p-$chains}

Notation: we drop boldface letters from now on. In general, a $p-$form $%
\beta $ is called \textit{closed} if its exterior derivative is equal to
zero,
\begin{equation*}
d\beta=0.
\end{equation*}
From this condition one can see that the closed form (the \textit{kernel} of
the exterior derivative operator $d$) is conserved quantity. Therefore,
closed $p-$forms possess certain invariant properties, physically
corresponding to the conservation laws (see e.g., \cite{Abraham}).

Also, a $p-$form $\beta$ that is an exterior derivative of some $(p-1)-$form
$\alpha$,
\begin{equation*}
\beta=d\alpha,
\end{equation*}
is called \textit{exact} (the \textit{image} of the exterior derivative
operator $d$). By Poincar\'e Lemma, exact forms prove to be closed
automatically,
\begin{equation*}
d\beta=d(d\alpha)=0.
\end{equation*}

Since $d^{2}=0$, \emph{every exact form is closed.} The converse
is only partially true (Poincar\'{e} Lemma): every closed form is
\textit{locally exact}. This means that given a closed $p-$form on
a smooth manifold $M$ (see Appendix), $\alpha \in \Omega ^{p}(M)$
on an open set $U\subset M$, any point $m\in U$ has a neighborhood
on which there exists a $(pk-1)-$form $\beta \in \Omega ^{p-1}(U)$
such that $d\beta =\alpha |_{U}.$

The Poincar\'{e} lemma is a generalization and unification of two
well--known facts in vector calculus:\newline
(i) If $\limfunc{curl}F=0$, then locally $F=\limfunc{grad}f$; ~ and ~ (ii)
If $\limfunc{div}F=0$, then locally $F=\limfunc{curl}G$.

\textit{Poincar\'{e} lemma} for contractible manifolds: Any closed form on a
smoothly contractible manifold is exact.

A \textit{cycle} is a chain $C\in \mathcal{C}_{p}(M)$ such that $\partial
C=0 $. A \textit{boundary} is a chain $C$ such that $C=\partial B,$ for any
other chain $B\in \mathcal{C}_{p}(M)$. Similarly, a \textit{cocycle} (i.e.,
a \textit{closed form}) is a cochain $\omega $ such that $d\omega =0$. A
\textit{coboundary} (i.e., an \textit{exact form}) is a cochain $\omega $
such that $\omega =d\theta ,$ for any other cochain $\theta $. All exact
forms are closed ($\omega =d\theta \Rightarrow d\omega =0$) and all
boundaries are cycles ($C=\partial B\Rightarrow \partial C=0$). Converse is
true only for smooth contractible manifolds, by Poincar\'{e} Lemma.

\subsection{Topological duality of $p-$forms and $p-$chains}

Duality of $p-$forms and $p-$chains on a smooth manifold $M$ is based on the
\begin{equation*}
\text{Period:}\ \ \int_{C}\omega :=\left\langle C,\omega \right\rangle ,
\end{equation*}%
where $C$ is a cycle, $\omega $ is a cocycle, while $\left\langle
C,\omega \right\rangle =\omega (C)$ is their inner product
$\left\langle C,\omega \right\rangle :\Omega ^{p}(M)\times
\mathcal{C}_{p}(M)\rightarrow \mathbb{R}$, a bilinear
non-degenerate functional (pairing) on $M$ \cite{Bruhat}. From
Poincar\'{e} Lemma, a closed $p-$form $\omega $ is exact iff
$\left\langle C,\omega \right\rangle =0$.

The fundamental topological duality is based on the Stokes theorem,
\begin{equation*}
\int_{\partial C}\omega =\int_{C}d\omega \qquad \mathrm{or}\qquad
\left\langle \partial C,\omega \right\rangle =\left\langle C,d\omega
\right\rangle ,
\end{equation*}%
where $\partial C$ is the boundary of the $p-$chain $C$ oriented coherently
with $C$ on $M$. While the \textit{boundary operator} $\partial $ is a
global operator, the coboundary operator $d$ is local, and thus more
suitable for applications. The main property of the exterior differential,
\begin{equation*}
d\circ d\equiv d^{2}=0 \quad \Longrightarrow \quad \partial \circ \partial
\equiv \partial ^{2}=0,\qquad (\text{and converse}),
\end{equation*}%
\noindent can be easily proved using the Stokes' theorem as
\begin{equation*}
0=\left\langle \partial ^{2}C,\omega \right\rangle =\left\langle \partial
C,d\omega \right\rangle =\left\langle C,d^{2}\omega \right\rangle =0.
\end{equation*}

\subsection{The de Rham cochain complex}

In the Euclidean 3D space $\Bbb{R}^{3}$ we have the following de Rham \emph{%
cochain complex}
\[
0\rightarrow \Omega ^{0}(\Bbb{R}^{3})\underset{\mathrm{grad}}{\stackrel{d}{%
\longrightarrow }}\Omega ^{1}(\Bbb{R}^{3})\underset{\mathrm{curl}}{%
\stackrel{d}{\longrightarrow }}\Omega ^{2}(\Bbb{R}^{3})\underset{\mathrm{%
div}}{\stackrel{d}{\longrightarrow }}\Omega
^{3}(\Bbb{R}^{3})\rightarrow 0.
\]
Using the \textit{closure property} for the exterior differential in $\Bbb{R}%
^{3},~d\circ d\equiv d^{2}=0$, we get the standard identities from
vector calculus
\[
\limfunc{curl}\cdot \limfunc{grad}=0\text{ \ \ \ \ \ \ and \ \ \ \ \ \ }%
\limfunc{div}\cdot \limfunc{curl}=0.
\]

As a duality, in $\Bbb{R}^{3}$ we have the following \emph{chain
complex}
\[
0\leftarrow \mathcal{C}_{0}(\Bbb{R}^{3}){\stackrel{\partial
}{\longleftarrow
}}\mathcal{C}_{1}(\Bbb{R}^{3}){\stackrel{\partial }{\longleftarrow }}%
\mathcal{C}_{2}(\Bbb{R}^{3}){\stackrel{\partial }{\longleftarrow }}\mathcal{C%
}_{3}(\Bbb{R}^{3})\leftarrow 0,
\]
(with the closure property $\partial\circ \partial\equiv
\partial^{2}=0$) which implies the following three boundaries:
\[
C_{1}\stackrel{\partial }{\mapsto }C_{0}=\partial (C_{1}),\qquad C_{2}%
\stackrel{\partial }{\mapsto }C_{1}=\partial (C_{2}),\qquad C_{3}\stackrel{%
\partial }{\mapsto }C_{2}=\partial (C_{3}),
\]
where $C_0\in{\cal C}_0$ is a 0--boundary (or, a point),
$C_1\in{\cal C}_1$ is a 1--boundary (or, a line), $C_2\in{\cal
C}_2$ is a 2--boundary (or, a surface), and $C_3\in{\cal C}_3$ is
a 3--boundary (or, a hypersurface). Similarly, the de Rham complex
implies the following three coboundaries:
\[
C^{0}\stackrel{d }{\mapsto }C^{1}=d (C^{0}),\qquad
C^{1}\stackrel{d }{\mapsto }C^{2}=d (C^{1}),\qquad
C^{2}\stackrel{d }{\mapsto }C^{3}=d (C^{2}),
\]
where $C^0\in\Omega^0$ is 0--form (or, a function),
$C^1\in\Omega^1$ is a 1--form, $C^2\in\Omega^2$ is a 2--form, and
$C^3\in\Omega^3$ is a 3--form.

In general, on a smooth $n$D manifold $M$ we have the following de
Rham cochain complex \cite{De Rham}
\begin{equation*}
0\rightarrow \Omega ^{0}(M)\overset{d}{\longrightarrow }\Omega ^{1}(M)%
\overset{d}{\longrightarrow }\Omega ^{2}(M)\overset{d}{\longrightarrow }%
\Omega ^{3}(M)\overset{d}{\longrightarrow }\cdot \cdot \cdot \overset{d}{%
\longrightarrow }\Omega ^{n}(M)\rightarrow 0,
\end{equation*}%
satisfying the closure property on $M,~d\circ d\equiv d^{2}=0$.
\begin{figure}[tbh]
\centerline{\includegraphics[width=13cm]{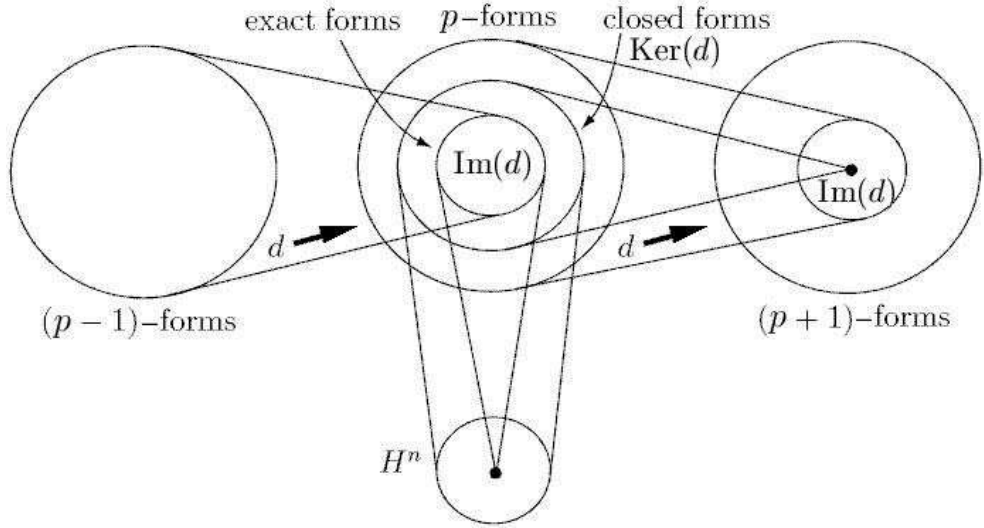}} \caption{A
small portion of the de Rham cochain complex, showing a
homomorphism of cohomology groups.} \label{Cohomology}
\end{figure}

\subsection{(Co)homology on a smooth manifold $M$}

The subspace of all closed $p-$forms (cocycles) on a smooth manifold $M$ is
the kernel $\limfunc{Ker}(d)$ of the de Rham $d-$homomorphism (see Figure %
\ref{Cohomology}), denoted by $Z^{p}(M)\subset \Omega ^{p}(M)$, and the
sub-subspace of all exact $p-$forms (coboundaries) on $M$ is the image $%
\limfunc{Im}(d)$ of the de Rham homomorphism denoted by $B^{p}(M)\subset
Z^{p}(M)$. The \textit{quotient space}
\begin{equation*}
H^{p}(M):=\frac{Z^{p}(M)}{B^{p}{M}}=\frac{\limfunc{Ker}\left( d:\Omega
^{p}(M)\rightarrow \Omega ^{p+1}(M)\right) }{\limfunc{Im}\left( d:\Omega
^{p-1}(M)\rightarrow \Omega ^{p}(M)\right) },
\end{equation*}%
is called the $p$th \textit{cohomology group} of a manifold $M$. It is a
topological invariant of a manifold. Two $p-$cocycles $\alpha $,$\beta \in
\Omega ^{p}(M)$ are \emph{homologous}, or belong to the same \textit{%
cohomology class} $[\alpha ]\in H^{p}(M)$, if they differ by a $(p-1)-$%
coboundary $\alpha -\beta =d\theta \in \Omega ^{p-1}(M)$.

Similarly, the subspace of all $p-$cycles on a smooth manifold $M$ is the
kernel $\limfunc{Ker}(\partial )$ of the $\partial -$homomorphism, denoted
by $Z_{p}(M)\subset \mathcal{C}_{p}(M)$, and the sub-subspace of all $p-$%
boundaries on $M$ is the image $\limfunc{Im}(\partial )$ of the $\partial -$%
homomorphism, denoted by $B_{p}(M)\subset \mathcal{C}_{p}(M)$. Two $p-$%
cycles $C_{1}$,$C_{2}\in \mathcal{C}_{p}$ are \emph{homologous}, if they
differ by a $(p-1)-$boundary $C_{1}-C_{2}=\partial B\in \mathcal{C}_{p-1}(M)$%
. Then $C_{1}$ and $C_{2}$ belong to the same \textit{homology class} $%
[C]\in H_{p}(M)$, which is defined as
\begin{equation*}
H_{p}(M):=\frac{Z_{p}(M)}{B_{p}(M)}=\frac{\limfunc{Ker}(\partial :\mathcal{C}%
_{p}(M)\rightarrow \mathcal{C}_{p-1}(M))}{\limfunc{Im}(\partial :\mathcal{C}%
_{p+1}(M)\rightarrow \mathcal{C}_{p}(M))},
\end{equation*}%
where $Z_{p}$ is the vector space of cycles and $B_{p}\subset Z_{p}$ is the
vector space of boundaries on $M$.

\subsection{A cochain complex spanning the space-time manifold}

Consider a small portion of the de Rham cochain complex of Figure \ref%
{Cohomology} spanning a space-time 4--manifold $M$,
\begin{equation*}
\Omega ^{p-1}(M)\overset{d_{p-1}}{\longrightarrow }\Omega ^{p}(M)\overset{%
d_{p}}{\longrightarrow }\Omega ^{p+1}(M)
\end{equation*}%
As we have seen above, cohomology classifies topological spaces by comparing
two subspaces of ~$\Omega ^{p}$:~ (i) the space of $p-$cocycles, $Z^{p}(M):=%
\func{Ker}d_{p}$, and~ (ii) the space of $p-$coboundaries, $B^{p}(M):=\func{%
Im}d_{p-1}$. Thus, for the cochain complex of any space-time 4--manifold we
have,
\begin{equation*}
d^{2}=0\quad \Longrightarrow \quad B^{p}(M)\subset Z^{p}(M),
\end{equation*}%
that is, every $p-$coboundary is a $p-$cocycle. Whether the converse of this
statement is true, according to Poincar\'{e} Lemma, depends on the
particular topology of a space-time 4--manifold. If every $p-$cocycle is a $%
p-$coboundary, so that $B^{p}$ and $Z^{p}$ are equal, then the cochain
complex is exact at $\Omega ^{p}(M)$. In topologically interesting regions
of a space-time manifld $M$, exactness may fail, and we measure the failure
of exactness by taking the $p$th cohomology group
\begin{equation*}
H^{p}(M):=Z^{p}(M)/B^{p}(M).
\end{equation*}

\subsection{Hodge star, codifferential and Laplacian}

The \textit{Hodge star} operator $\star :\Omega ^{p}(M)\rightarrow \Omega
^{n-p}(M)$ on a smooth manifold $M$ is defined as \cite{Hodge}
\begin{equation*}
\alpha \wedge \star \,\beta =\beta \wedge \star \,\alpha =\left\langle
\alpha ,\beta \right\rangle \mu ,\qquad \text{for }\alpha ,\beta \in \Omega
^{p}(M).
\end{equation*}%
The $\star $ operator depends on the Riemannian metric on $M$ and also on
the orientation (reversing orientation will change the sign). The \textit{%
volume form} $\mu $ is defined in local coordinates on $M$ as
\begin{equation*}
\mu =\text{vol}=\star (1)=\sqrt{\det (g_{ij})}~dx^{1}\wedge ...\wedge dx^{n},
\end{equation*}%
and the total volume on $M$ is given by
\begin{equation*}
\limfunc{vol}(M)=\int_{M}\star (1).
\end{equation*}%
For example, in the 4D electrodynamics, the dual 2--form \emph{Maxwell} $%
\star F$ satisfies the electric Maxwell equation with the source \cite{MTW},
\begin{equation*}
\text{Dual Bianchi identity}:\quad\mathbf{d}\mathbf{\star F}=\mathbf{\star J}%
,
\end{equation*}%
where $\mathbf{\star J}$ is the 3--form dual to the charge--current 1--form $%
\mathbf{J}$.

For any to $p-$forms $\alpha ,\beta \in \Omega ^{p}(M)$ with compact
support, we define the (bilinear and positive--definite) Hodge $L^{2}-$inner
product as
\begin{equation}
(\alpha ,\beta )=\int_{M}\langle \alpha ,\beta \rangle \star
(1)=\int_{M}\alpha \wedge \star \beta .  \label{L2}
\end{equation}%
We can extend the product $(\cdot ,\cdot )$ to $L^{2}(\Omega ^{p}(M))$; it
remains bilinear and positive--definite, because as usual in the definition
of $L^{2}$, functions that differ only on a set of measure zero are
identified.

Now, the Hodge dual to the exterior derivative $d:\Omega ^{p}(M)\rightarrow
\Omega ^{p+1}(M)$ on a smooth manifold $M$ is the \emph{codifferential} $%
\delta $, a linear map $\delta :\Omega ^{p}(M)\rightarrow \Omega ^{p-1}(M)$,
which is a generalization of the divergence, defined by \cite{Hodge}
\begin{equation*}
\delta =\star \,d\star (-1)^{n(p+1)+1},\qquad \text{so that\qquad }d=\star
\,\delta \star (-1)^{np}.
\end{equation*}%
Applied to any $p-$form $\omega \in \Omega ^{p}(M)$, the codifferential $%
\delta $ gives
\begin{equation*}
\delta \omega =(-1)^{n(p+1)+1}\star d\star \omega .
\end{equation*}%
If $\omega =f$ is a $0-$form, or function, then $\delta f=0$.

The codifferential $\delta $ can be coupled with the exterior derivative $d$
to construct the \emph{Hodge Laplacian} $\Delta :$ $\Omega
^{p}(M)\rightarrow \Omega ^{p}(M),$ a harmonic generalization of the
Laplace--Beltrami differential operator, given by\footnote{%
The difference $d-\delta ={\partial _{D}}$ is called the \textit{Dirac
operator}. Its square $\partial _{D}^{2}$\ equals the Hodge Laplacian $%
\Delta $.}
\begin{equation*}
\Delta =\delta d+d\delta .
\end{equation*}

The Hodge codifferential $\delta $ satisfies the following set of rules:

\begin{itemize}
\item $\delta \delta =\delta ^{2}=0,$ \ the same as $\ dd=d^{2}=0;$

\item $\delta \Delta =\Delta \delta ,$ \ the same as $\ d\Delta =\Delta d;$

\item $\delta \star =(-1)^{p+1}\star d$; \ $\star \,\delta =(-1)^{p}\star d$;

\item $d\delta \star =\star \,\delta d$; \ $\star \,d\delta =\delta d\star$;
\ $\star \,\Delta =\Delta \star .$
\end{itemize}

\section{Classical Gauge Electrodynamics in 4D}

Recall that a \textit{gauge theory} is a theory that admits a symmetry with
a local parameter. For example, in every quantum theory the global phase of
the wave $\psi -$function is arbitrary and does not represent something
physical. Consequently, the theory is invariant under a global change of
phases (adding a constant to the phase of all wave functions, everywhere);
this is a global symmetry. In quantum electrodynamics, the theory is also
invariant under a local change of phase, that is, one may shift the phase of
all wave functions so that the shift may be different at every point in
space-time. This is a local symmetry. However, in order for a well--defined
derivative operator to exist, one must introduce a new field, the \textit{%
gauge field}, which also transforms in order for the local change of
variables (the phase in our example) not to affect the derivative. In
quantum electrodynamics this gauge field is the electromagnetic potential
1--form $A$, in components given by
\begin{equation*}
A=A_{\mu }dx^{\mu },\quad \text{such that}\quad A_{\mathrm{new}}=A_{\mathrm{%
old}}+df,
\end{equation*}
(where $f$ is an arbitrary scalar function) -- leaves the electromagnetic
field 2--form $F=dA$ unchanged. This change $df$ of local gauge of variable $%
A$ is termed \textit{gauge transformation}. In quantum field theory the
excitations of fields represent particles. The particle associated with
excitations of the gauge field is the \emph{gauge boson}. All the
fundamental interactions in nature are described by gauge theories. In
particular, in quantum electrodynamics, whose gauge transformation is a
local change of phase, the gauge group is the circle group $U(1)$
(consisting of all complex numbers with absolute value $1$), and the gauge
boson is the photon (see e.g., \cite{Frampton}).

In the 4D space-time electrodynamics, the 1--form \textit{electric current
density} $J$ has the components $J_{\mu }=(\rho ,\mathbf{j})=(\rho
,j_{x},j_{y},j_{z})$ (where $\rho $ is the charge density), the 2--form
\textit{Faraday} $F$ is given in components of electric field $\mathbf{E}$
and magnetic field $\mathbf{B}$ by
\begin{equation*}
F_{\mu \nu }=\left(
\begin{array}{cccc}
0 & E_{x} & E_{y} & E_{z} \\
-E_{x} & 0 & -B_{z} & B_{y} \\
-E_{y} & B_{z} & 0 & -B_{x} \\
-E_{z} & -B_{y} & B_{x} & 0%
\end{array}%
\right) ,\qquad \text{with\qquad }F_{\nu \mu }=-F_{\mu \nu }
\end{equation*}%
while its dual 2--form \textit{Maxwell} $\star F$ has the following
components
\begin{equation*}
\star F_{\mu \nu }=\left(
\begin{array}{cccc}
0 & -B_{x} & -B_{y} & -B_{z} \\
B_{x} & 0 & -E_{z} & E_{y} \\
B_{y} & E_{z} & 0 & -E_{x} \\
B_{z} & -E_{y} & B_{x} & 0%
\end{array}%
\right) ,\qquad \text{with\qquad }\star F_{\nu \mu }=-\star F_{\mu \nu }.
\end{equation*}

\subsection{Maxwell's equations}

The \textit{gauge field} in classical electrodynamics, in local coordinates
given as an electromagnetic potential 1--form
\begin{equation*}
A=A_{\mu }dx^{\mu }=A_{\mu }dx^{\mu }+df,\qquad (f=~\text{arbitrary scalar
field}),
\end{equation*}%
is globally a \textit{connection} on a $U(1)-$bundle\footnote{%
Recall that in the 19th Century, Maxwell unified Faraday's electric and
magnetic fields. Maxwell's theory led to Einstein's special relativity where
this unification becomes a spin-off of the unification of space end time in
the form of the \textit{Faraday tensor} \cite{MTW}
\begin{equation*}
F=E\wedge dt+B,
\end{equation*}%
where $F$ is electromagnetic $2-$form on space-time, $E$ is electric $1-$%
form on space, and $B$ is magnetic $2-$form on space. Gauge theory considers
$F$ as secondary object to a connection--potential $1-$form $A$. This makes
half of the Maxwell equations into tautologies \cite{BaezGauge}, i.e.,
\begin{equation*}
F=dA\quad \Longrightarrow \quad dF=0\quad :\quad \text{Bianchi identity},
\end{equation*}%
but does not imply the second half of Maxwell's equations,
\begin{equation*}
\delta F=-4\pi J\quad :\quad \text{dual Bianchi identity}.
\end{equation*}%
To understand the deeper meaning of the connection--potential $1-$form $A$,
we can integrate it along a path $\gamma $ in space-time, ~$x\cone{\gamma}y$%
.~ Classically, the integral $\int_{\gamma }A$ represents an \emph{action}
for a charged point particle to move along the path $\gamma $.
Quantum--mechanically, $\exp \left( \mathrm{i}\int_{\gamma }A\right) $
represents a \emph{phase} (within the unitary Lie group $U(1)$) by which the
particle's wave--function changes as it moves along the path $\gamma $, so $%
A $ is a $U(1)-$connection.
\par
In other words, Maxwell's equations can be formulated using
complex line bundles, or principal bundles with fibre $U(1)$. The
connection $\nabla $ on the line bundle has a curvature $F=\nabla
^{2}$ which is a 2--form that automatically satisfies $dF=0$\ and
can be interpreted as a field--strength. If the line bundle is
trivial with flat reference connection $d,$ we can write $\nabla
=d+A$ \ and $F=dA$ with $A$ the 1--form composed of the electric
potential and the magnetic vector potential.} (see Appendix). The
corresponding electromagnetic field, locally the 2--form
\begin{eqnarray*}
F &=&dA,\qquad \text{in components given by} \\
F &=&\frac{1}{2}F_{\mu \nu }\,dx^{\mu }\wedge dx^{\nu },\qquad \text{with \
\ \ }F_{\mu \nu }=\partial _{\nu }A_{\mu }-\partial _{\mu }A_{\nu }
\end{eqnarray*}%
is globally the \textit{curvature} of the connection $A$\footnote{%
The only thing that matters here is the \emph{difference} $\alpha $ \emph{%
between two paths} $\gamma _{1}$ and $\gamma _{2}$ \emph{in the action} $%
\int_{\gamma }A$ \cite{BaezGauge}, which is a 2--morphism (see e.g., \cite%
{GaneshSprBig,GaneshADG})
\begin{equation*}
x\ctwodbl{\gamma_1}{\gamma_2}{\alpha}y
\end{equation*}%
} under the gauge--covariant derivative,
\begin{equation*}
D_{\mu }=\partial _{\mu }-{\rm i}eA_{\mu },
\end{equation*}%
where ${\rm i}=\sqrt{-1}$ and $e$ is the charge coupling constant.\footnote{%
If a gauge transformation is given by%
\begin{equation*}
\psi \mapsto \mathrm{e}^{i\Lambda }\psi
\end{equation*}%
and for the gauge potential%
\begin{equation*}
A_{\mu }\mapsto A_{\mu }+\frac{1}{e}(\partial _{\mu }\Lambda ),
\end{equation*}%
then the gauge--covariant derivative,
\begin{equation*}
D_{\mu }=\partial _{\mu }-ieA_{\mu }
\end{equation*}%
transforms as%
\begin{equation*}
D_{\mu }\mapsto \partial _{\mu }-{\rm i}eA_{\mu }-i(\partial _{\mu
}\Lambda )
\end{equation*}%
and $D_{\mu }\psi $ transforms as%
\begin{equation*}
D_{\mu }\mapsto \partial _{\mu }-ieA_{\mu }-i(\partial _{\mu }\Lambda ).
\end{equation*}%
}

Classical electrodynamics is governed by the \emph{Maxwell equations}, which
in exterior formulation read
\begin{eqnarray*}
dF &=&0,\qquad \delta F=-4\pi J,\qquad \text{or in components,} \\
F_{[\mu \nu ,\eta ]} &=&0,\qquad F_{\mu \nu },^{\mu }=-4\pi J_{\mu },
\end{eqnarray*}%
where comma denotes the partial derivative and the 1--form of electric
current $J=J_{\mu }dx^{\mu }$ is conserved, by the electrical \textit{%
continuity equation},
\begin{equation*}
\delta J=0,\qquad \text{or in components,\qquad }J_{\mu },^{\mu }=0.
\end{equation*}

The first, sourceless Maxwell equation, $dF=0$, gives vector magnetostatics
and magnetodynamics,
\begin{eqnarray*}
\text{Magnetic Gauss' law} &:&\func{div}\mathbf{B}=0,\qquad \\
\text{Faraday's law} &\text{:}&\partial _{t}\mathbf{B}+\func{curl}\mathbf{E}%
=0.
\end{eqnarray*}%
The second Maxwell equation with source, $\delta F=J$, gives vector
electrostatics and electrodynamics,
\begin{eqnarray*}
\text{Electric Gauss' law} &:&\func{div}\mathbf{E}=4\pi \rho ,\qquad \\
\text{Amp\`{e}re's law} &:&\partial _{t}\mathbf{E}-\func{curl}\mathbf{B}%
=-4\pi \mathbf{j}.
\end{eqnarray*}
Maxwell's equations are generally applied to macroscopic averages of the
fields, which vary wildly on a microscopic scale in the vicinity of
individual atoms, where they undergo quantum effects as well (see below).

\subsection{Electrodynamic action}

The standard \textit{Lagrangian} for the free electromagnetic field, $F=dA$,
is given by \cite{GaneshSprBig,GaneshADG,QuLeap}
\begin{equation*}
\mathcal{L}(A)=\frac{1}{2}(F\wedge \star \,F),
\end{equation*}%
with the corresponding action%
\begin{equation*}
S(A)=\frac{1}{2}\int F\wedge \star \,F.
\end{equation*}%
Using the Hodge $L^{2}-$inner product (\ref{L2}), we can rewrite this action
as%
\begin{equation}
S(A)=\frac{1}{2}(F,F).  \label{act}
\end{equation}

\subsection{2D electrodynamics}

One of the reasons that gauge electrodynamics in 2D is especially simple is
that the electromagnetic field $F=dA$, being a 2--form, can be written
simply as a scalar field $\varphi =\varphi (x,t)$ times the volume form \cite%
{Witten}
\begin{equation*}
F=\varphi \;\mathrm{vol}.
\end{equation*}%
This means that the only Hodge duals we need to consider are the trivial
ones,
\begin{equation*}
\star ~\mathrm{vol}=1\qquad ~\text{and}\qquad \star 1=\mathrm{vol},
\end{equation*}%
so the 2D Lagrangian for vacuum electrodynamics is given by \cite{Wise}
\begin{equation*}
L=\frac{1}{2}(\varphi \;\mathrm{vol})\wedge \star \,(\varphi \;\mathrm{vol})=%
\frac{1}{2}\varphi ^{2}\;\mathrm{vol}\wedge 1=\frac{\varphi ^{2}}{2}\mathrm{%
vol}.
\end{equation*}

\section{Quantum Electrodynamics in 4D}

Quantum electrodynamics (QED) is a relativistic quantum field
theory of electrodynamics. QED was developed by four Nobel
Laureates, P. Dirac, R. Feynman, J. Schwinger and S. Tomonaga, and
F. Dyson, between 1920s and 1950s. It describes some aspects of
how electrons, positrons and photons interact. QED mathematically
describes all phenomena involving electrically charged particles
interacting by means of exchange of photons.

\subsection{Dirac QED}

The \textit{Dirac equation} for a particle with mass $m$ (in
natural units) reads (see, e.g., \cite{QuLeap})
\begin{equation}
({\rm i}\gamma ^{\mu }\partial _{\mu }-m)\psi =0,\qquad (\mu
=0,1,2,3) \label{DiracEq}
\end{equation}%
where ${\rm i}=\sqrt{-1}$ $\psi (x)$ is a 4--component spinor\footnote{%
The most convenient definitions for the 2--spinors, like the Dirac
spinor, are:
\par
$\phi ^{1}=%
\begin{bmatrix}
1 \\
0%
\end{bmatrix}%
,\,\phi ^{2}=%
\begin{bmatrix}
0 \\
1%
\end{bmatrix}%
\,$ and $\chi ^{1}=%
\begin{bmatrix}
0 \\
1%
\end{bmatrix}%
,\,\chi ^{2}=%
\begin{bmatrix}
1 \\
0%
\end{bmatrix}%
\,.$} wave--function, the so--called Dirac spinor, while $\gamma
^{\mu }$ are $4\times 4$ \textit{Dirac }$\gamma
-$\textit{matrices},
\begin{eqnarray*}
&\gamma ^{0}=%
\begin{pmatrix}
1 & 0 & 0 & 0 \\
0 & 1 & 0 & 0 \\
0 & 0 & -1 & 0 \\
0 & 0 & 0 & -1%
\end{pmatrix}%
,\qquad &\gamma ^{1}\!=\!%
\begin{pmatrix}
0 & 0 & 0 & 1 \\
0 & 0 & 1 & 0 \\
0 & -1 & 0 & 0 \\
-1 & 0 & 0 & 0%
\end{pmatrix}%
, \\
&\gamma ^{2}\!=\!%
\begin{pmatrix}
0 & 0 & 0 & -i \\
0 & 0 & i & 0 \\
0 & i & 0 & 0 \\
-i & 0 & 0 & 0%
\end{pmatrix}%
,\qquad &\gamma ^{3}\!=\!%
\begin{pmatrix}
0 & 0 & 1 & 0 \\
0 & 0 & 0 & -1 \\
-1 & 0 & 0 & 0 \\
0 & 1 & 0 & 0%
\end{pmatrix}%
.
\end{eqnarray*}%
They obey the \textit{anticommutation relations}
\begin{equation*}
\{\gamma ^{\mu },\gamma ^{\nu }\}=\gamma ^{\mu }\gamma ^{\nu
}+\gamma ^{\nu }\gamma ^{\mu }=2g^{\mu \nu },
\end{equation*}%
where $g_{\mu \nu }$ is the metric tensor.

Dirac's $\gamma -$matrices are conventionally derived as%
\begin{equation*}
\gamma ^{k}=%
\begin{pmatrix}
0 & \sigma ^{k} \\
-\sigma ^{k} & 0%
\end{pmatrix}%
,\qquad (k=1,2,3)
\end{equation*}%
where $\sigma ^{k}$ are \textit{Pauli }$\sigma -$\textit{matrices}\footnote{%
In quantum mechanics, each Pauli matrix represents an observable
describing the spin of a spin ${\frac12} $ particle in the three
spatial directions. Also, $i\sigma _{j}$ are the generators of
rotation acting on non-relativistic particles with spin ${\frac12}
$. The state of the particles are represented as two--component
spinors.
\par
In \emph{quantum information}, single--qubit quantum gates are
$2\times 2$ unitary matrices. The Pauli matrices are some of the
most important single--qubit operations.} (a set of $2\times 2$
complex Hermitian and
unitary matrices), defined as%
\begin{equation*}
\sigma _{1}=\sigma _{x}=%
\begin{pmatrix}
0 & 1 \\
1 & 0%
\end{pmatrix}%
,\qquad \sigma _{2}=\sigma _{y}=%
\begin{pmatrix}
0 & -i \\
i & 0%
\end{pmatrix}%
,\qquad \sigma _{3}=\sigma _{z}=%
\begin{pmatrix}
1 & 0 \\
0 & -1%
\end{pmatrix}%
,
\end{equation*}%
obeying both the commutation and anticommutation relations%
\begin{equation*}
\lbrack \sigma _{i},\sigma _{j}] = 2i\,\varepsilon _{ijk}\,\sigma
_{k},\qquad \{\sigma _{i},\sigma _{j}\} =2\delta _{ij}\cdot I,
\end{equation*}
where $\varepsilon _{ijk}$\ is the Levi--Civita symbol, $\delta
_{ij}$ is the Kronecker delta, and $I$ is the identity matrix.

Now, the Lorentz--invariant form of the Dirac equation
(\ref{DiracEq}) for an electron with a charge $e$ and mass
$m_{\mathrm{e}}$, moving with a 4--momentum 1--form $p=p_{\mu
}dx^{\mu }$ in a classical electromagnetic
field defined by 1--form $A=A_{\mu }dx^{\mu }$, reads (see, e.g., \cite%
{Drake,QuLeap}):
\begin{equation}
\left\{{\rm i}\gamma ^{\mu }\left[ p_{\mu }-eA_{\mu }\right] -m_{\mathrm{e}%
}\right\} \psi (x)=0,  \label{DiracCov}
\end{equation}%
and is called the \textit{covariant Dirac equation}.

The formal QED Lagrangian (density) includes three terms,%
\begin{equation}
\mathcal{L}(x)=\mathcal{L}_{\mathrm{em}}(x)+\mathcal{L}_{\mathrm{int}}(x)+%
\mathcal{L}_{\mathrm{e-p}}(x),  \label{QEDlagrangian}
\end{equation}%
related respectively to the free electromagnetic field 2--form
$F=F_{\mu \nu }dx^{\mu }\wedge dx^{\nu }$, the electron--positron
field (in the presence of the external vector potential 1--form
$A_{\mu }^{\mathrm{ext}})$, and the interaction field (dependent
on the charge--current 1--form $J=J_{\mu
}dx^{\mu }$). The free electromagnetic field Lagrangian in (\ref%
{QEDlagrangian}) has the standard electrodynamic form%
\begin{equation*}
\mathcal{L}_{\mathrm{em}}(x)=-\frac{1}{4}F^{\mu \nu }F_{\mu \nu },
\end{equation*}%
where the electromagnetic fields are expressible in terms of
components of the potential 1--form\newline
$A=A_{\mu }dx^{\mu }$ by%
\begin{equation*}
F_{\mu \nu }=\partial _{\mu }A_{\nu }^{\mathrm{tot}}-\partial
_{\nu }A_{\mu
}^{\mathrm{tot}},\quad \text{with\quad }A_{\mu }^{\mathrm{tot}}=A_{\mu }^{%
\mathrm{ext}}+A_{\mu }.
\end{equation*}

The electron-positron field Lagrangian is given by Dirac's equation (\ref%
{DiracCov}) as
\begin{equation*}
\mathcal{L}_{\mathrm{e-p}}(x)=\bar{\psi}(x)\left\{{\rm i}\gamma
^{\mu }\left[ p_{\mu }-eA_{\mu }^{\mathrm{ext}}\right]
-m_{\mathrm{e}}\right\} \psi (x),
\end{equation*}%
where $\bar{\psi}(x)$ is the Dirac adjoint spinor wave function.

The interaction field Lagrangian%
\begin{equation*}
\mathcal{L}_{\mathrm{int}}(x)=-J^{\mu }A_{\mu },
\end{equation*}%
accounts for the interaction between the uncoupled electrons and
the radiation field.

The field equations deduced from (\ref{QEDlagrangian}) read%
\begin{eqnarray}
\left\{{\rm i}\gamma ^{\mu }\left[ p_{\mu }-eA_{\mu }^{\mathrm{ext}}\right] -m_{%
\mathrm{e}}\right\} \psi (x) &=&\gamma ^{\mu }\psi (x)A_{\mu },  \notag \\
\partial ^{\mu }F_{\mu \nu } &=&J_{\nu }.  \label{QEDformal}
\end{eqnarray}%
The formal QED requires the solution of the system (\ref{QEDformal}) when $%
A^{\mu }(x),$ $\psi (x)$ and $\bar{\psi}(x)$ are quantized fields.

\subsection{Feynman QED}

In Feynman's form of quantum electrodynamics
\cite{FeynQED,GaneshADG,QuLeap}, the process of quantization is
transparent -- it is performed using his path integral formalism,
resulting in the quantum--field transition  given (in natural
units) by:
\begin{equation} \langle out_AA|in_A\rangle=\int\mathcal{D}[A]\, {\mathrm
e}^{{\mathrm i} S(A)}\quad \underrightarrow{Wick}\quad
\int\mathcal{D}[A]\, {\mathrm e}^{- S(A)},\qquad \label{field}
\end{equation}
with action $S(A)$ given by (\ref{act}) in the exponent
(multiplied by the imaginary unit i). The Lebesgue integration in
(\ref{field}) is performed over electromagnetic fields $A_\mu$,
with the Lebesgue measure
\begin{equation*}
\mathcal{D}[A]=\lim_{N\to\infty}\prod_{s=1}^{N}dA^s_\mu \qquad
(\mu=1,..,4).
\end{equation*}
The transformation $\underrightarrow{Wick}$ denotes the so--called
Wick--rotation of the time variable $t$ to imaginary values
$t\mapsto \tau={\mathrm i} t$, thereby transforming the
complex-valued path integral into the real-valued (Euclidean) one.
The absolute square of the transition amplitude (\ref{field})
gives the transition probability density function, $P(A)=|\langle
out_A|in_A\rangle|^2$. Full discretization of (\ref{field})
ultimately gives the standard {thermodynamic partition function}
\begin{equation}
Z=\sum_j{\mathrm e}^{-E^j/T}, \label{partition}
\end{equation}
where $E^j$ is the energy eigenvalue, $T$ is the temperature
environmental control parameter, and the sum runs over all $j$
energy eigenstates. From (\ref{partition}), we can further
calculate all thermodynamical and statistical properties
\cite{FeynmanStat}, as for example, {transition entropy} $S =
k_B\ln Z$, etc.

The standard QED considers the path integral
\begin{eqnarray*}
Z[A] &=&\int \mathcal{D}[A]\,\mathrm{e}^{\mathrm{i}S[A]},\qquad \text{where
the action for the free e.-m. field is} \\
S[A] &=&\int d^{4}x\left[ -\frac{1}{4}(F_{\mu \nu })^{2}\right] =\frac{1}{2}%
\int d^{4}x\,A_{\mu }(x)\left( \partial ^{2}g^{\mu \nu }-\partial ^{\mu
}\partial ^{\nu }\right) A_{\nu }(x).
\end{eqnarray*}

$Z[A]$ is the path integral over each of the four spacetime components:
\begin{equation*}
\mathcal{D}[A]=\mathcal{D}[A]^{0}\mathcal{D}[A]^{1}\mathcal{D}[A]^{2}%
\mathcal{D}[A]^{3}.
\end{equation*}
This functional integral is badly divergent, due to gauge invariance. Recall
that $F_{\mu \nu },$ and hence $L,$ is invariant under a general gauge
transformation of the form
\begin{equation*}
A_{\mu }(x)\rightarrow A_{\mu }(x)+\frac{1}{e}\partial _{\mu }\alpha (x).
\end{equation*}
The troublesome modes are those for which \ $A_{\mu }(x)=\partial _{\mu
}\alpha (x),$ that is, those that are gauge--equivalent to \ $A_{\mu }(x)=0.$
The path integral is badly defined because we are redundantly integrating
over a continuous infinity of physically equivalent field configurations. To
fix this problem, we would like to isolate the interesting part of the path
integral, which counts each physical configuration only once. This can be
accomplished using the \emph{Faddeev--Popov trick} \cite{Faddeev}, which effectively adds a
term to the system Lagrangian and after which we get
\begin{equation*}
Z[A]=\int \mathcal{D}[A]\exp \left[ \mathrm{i}\int_{-T}^{T}d^{4}x\,\left[
\mathcal{L}-\frac{1}{2\xi }(\partial ^{\mu }A_{\mu })^{2}\right] \right],
\end{equation*}
where $\xi $ is any finite constant.

Now, the {observable} $O$ in the {quantum field theory} is a
real--valued function of the gauge field: $O\colon
\mathcal{A}\rightarrow \mathbb{R},$ where $\mathcal{A}$ is the
{space of connections}, so we can try to calculate its expected
value using the electrodynamic action $S(A) $ (\ref{act}) as
\begin{equation*}
\left\langle \Omega |T\,\mathcal{O}(A)|\Omega \right\rangle =\frac{\int_{\mathcal{A}}\mathcal{D}[A]O(A)\,\mathrm{e}%
^{-S(A)}}{\int_{\mathcal{A}}\mathcal{D}[A]\,\mathrm{e}^{-S(A)}}.
\end{equation*}%
In certain cases, it might be difficult to define exactly what this means.
Even in the finite dimensional case, where $\mathcal{D}[A]$ is the {%
Lebesgue measure} on $\mathcal{A}$, when we try to calculate
$\langle O\rangle $ by the above formula we usually run into
serious problems. To see this, let us do a sample calculation of
the integral in the denominator, which is the {partition function}
\begin{equation}
Z:=\int_{\mathcal{A}}\mathcal{D}[A]\,\mathrm{e}^{-S(A)}
\label{partit}
\end{equation}%
against which all other expected values are normalized.

Thus, the Faddeev--Popov trick needs also to be applied to the formula for the two--point
correlation function
\begin{equation*}
\left\langle \Omega |T\,\mathcal{O}(A)|\Omega \right\rangle
=\lim_{T\rightarrow \infty (1-i\epsilon )}\frac{\int \mathcal{D}[A]\,%
\mathcal{O}(A)\exp \left[ \mathrm{i}\int_{-T}^{T}d^{4}x\,\mathcal{L}\right]
\mathcal{\,}}{\int \mathcal{D}[A]\exp \left[ \mathrm{i}\int_{-T}^{T}d^{4}x\,%
\mathcal{L}\right] },
\end{equation*}
which after Faddeev--Popov procedure becomes
\begin{equation*}
\left\langle \Omega |T\,\mathcal{O}(A)|\Omega \right\rangle
=\lim_{T\rightarrow \infty (1-i\epsilon )}\frac{\int \mathcal{D}[A]\,%
\mathcal{O}(A)\exp \left[ \mathrm{i}\int_{-T}^{T}d^{4}x\,\left[ \mathcal{L}-%
\frac{1}{2\xi }(\partial ^{\mu }A_{\mu })^{2}\right] \right] \mathcal{\,}}{%
\int \mathcal{D}[A]\exp \left[ \mathrm{i}\int_{-T}^{T}d^{4}x\,\left[
\mathcal{L}-\frac{1}{2\xi }(\partial ^{\mu }A_{\mu })^{2}\right] \right] }.
\end{equation*}

The space of gauge transformations, $\mathcal{G}$, is a group
which acts on the space $\mathcal{A}$ of connections by
\begin{equation*}
\mathcal{G}\times \mathcal{A}\rightarrow \mathcal{A},\qquad (\phi ,A)\mapsto
A+d\phi .
\end{equation*}%
A common way of eliminating divergences caused by gauge freedom is {%
gauge fixing }\cite{Wise}, which means choosing some method to
pick one representative of each gauge--equivalence class $[A]$ and
doing path integrals over these. There are problems with this
approach, however. In a general gauge theory, it might not even be
possible to fix a smooth, global gauge over which to integrate.
But even when we can do this, the arbitrary choice involved in
fixing a gauge is undesirable, philosophically. Gauge fixing
amounts to pretending a {quotient space} of $\mathcal{A}$, the
space of {connections modulo gauge transformations}, is a {%
subspace}. A better approach is to use the quotient space directly. Namely,
modding out by the action of $\mathcal{G}$ on $\mathcal{A}$ gives the
quotient space $\mathcal{A/G}$ consisting of gauge--equivalence classes of
connections, and we do path integrals like
\begin{equation*}
Z=\int_{\mathcal{A}/\mathcal{G}}d[A]\,\mathrm{e}^{-S([A])}.
\end{equation*}%
When gauge fixing works, path integrals over $\mathcal{A/G}$ give the same
results as gauge fixing. But integrating over $\mathcal{A/G}$ is more
general and involves no arbitrary choices.

Since gauge equivalent connections are regarded as physically
equivalent, the quotient space ${\mathcal{A}/\mathcal{G}}$ of
connections mod gauge transformations is sometimes called the
{physical configuration space} for vacuum electrodynamics. A
{physical observable} is then any
real--valued function on the physical configuration space, $O:{\mathcal{A}/%
\mathcal{G}}\rightarrow \mathbb{R},$ or equivalently, any
gauge--invariant function on the space $\mathcal{A}$ of
connections. In the case of noncompact gauge group even factoring
out all of the gauge freedom may be insufficient to regularize our
path integrals. In particular, there are certain topological
conditions our space-time must meet for this programme to give
convergent path integrals. As a consequence, in the case where the
gauge group is $\mathbb{R}$, we must sometimes take more drastic
measures in order to extract meaningful results, and this leads to
some interesting differences between the cases $G=\mathbb{R}$ and
$G=U(1)$ as the gauge group for electrodynamics. These differences
are related to the famous `Bohm--Aharonov effect' \cite{Wise}.

\subsection{Cohomology criterion for path--integral convergence}

To eliminate the divergences in the path integral (\ref{partit}) it is
necessary that we eliminate any gauge freedom from the space of connections $%
\mathcal{A}$. In gauge theory, the action $S(A)$ is invariant under the
group of gauge transformations $\mathcal{G}$. In case of a gauge group $%
\mathcal{G}=\mathbb{R}$, we have gauge transformations of the form
\begin{equation*}
A\mapsto A+d\varphi ,
\end{equation*}%
where $\varphi $ is an arbitrary $(p-1)-$cochain. In other words, two $p-$%
connections are {gauge equivalent} if they differ by a
$p-$coboundary $d\varphi \in B^{p}$. Before we saw that the
existence of gauge freedom implied path integrals such as the one
above must diverge \cite{Wise}. The standard procedure of
eliminating any gauge freedom from the space of connections
$\mathcal{A}$ is to pass to the quotient space
\begin{equation*}
\mathcal{A}/\mathcal{G}=C^{p}/B^{p},
\end{equation*}
that is, $p-$connections modulo gauge transformations, effectively
declaring {gauge equivalent} $p-$connections to be equal. In
particuclar, the criterion in $\mathbb{R}$
electrodynamics for path integrals over the physical configuration space $%
\mathcal{A/G}$ to converge is that the first cohomology be trivial \cite%
{Wise}:
\begin{equation*}
\left(
\begin{array}{c}
Z=\int_{\mathcal{A/G}}\mathcal{D}[A]\,\mathrm{e}^{-S(A)} \\
\text{converges for $\mathbb{R}$ electrodynamics}%
\end{array}%
\right) \iff H^{1}=0.
\end{equation*}

\section{Electro--Muscular Stimulation}

In this section we develop covariant biophysics of
electro--muscular stimulation, as an externally induced generator
of the \textit{covariant muscular force},
\begin{equation}
F_{i}=m_{ij}\,a^{i}, \label{cov}
\end{equation} where $F_{i}$ are covariant components
of the muscular force (torque) 1--form $F=F_i(t)\,dx^i$, $m_{ij}$
are the covariant components of the musculo-skeletal
metric/inertia tensor $m=m_{ij}\;dx^i\otimes dx^j$, while
$a^{i}=a^{i}(t)\,\partial_{x^i}$ are contravariant components of
the angular acceleration vector (see
\cite{GaneshSprSml,GaneshSprBig} for technical details). The
so--called \textit{functional electrical stimulation} (FES) of
human skeletal muscles is used in rehabilitation and in medical
orthotics to externally stimulate the muscles with damaged neural
control. However, the repetitive use of electro--muscular
stimulation, besides functional, causes also structural changes in
the stimulated muscles, giving the physiological effect of
muscular training.

\begin{figure}[h]
\centerline{\includegraphics[width=12cm]{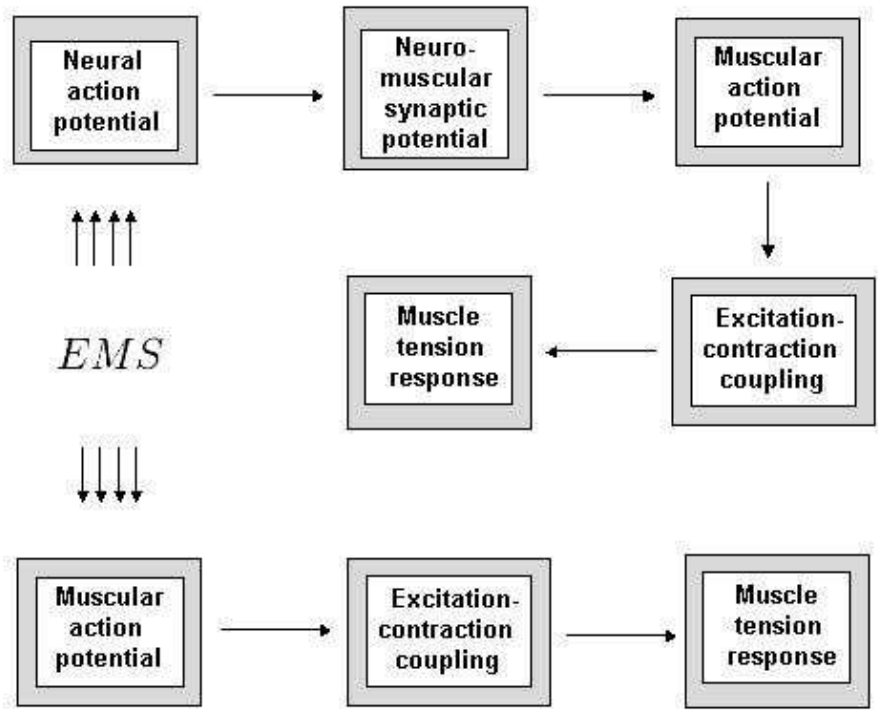}} \caption{Schematic
of electro--muscular stimulation ($EMS$).}\label{FES}
\end{figure}

The use of low and very low frequency impulses in the body,
delivered through electrodes, is known as transcutaneous
stimulation of the nerves, electro--acupuncture and
electro--stimulation. Here, an electromagnetic field accompanies
the passage of the electric current through the conductive wire.
This is generally known as the term `electromagnetic
therapy'.\footnote{In the original sense acupuncture meant the
inserting of needles in specific regions of the body.
Electro--acupuncture supplies the body with low--volt impulses
through the medium of surface electrodes to specific body regions
or by non specific electrodes. Transcutaneous electric stimulation
of the nerves (TENS) has for years been a well known procedure in
conventional medicine. The impulses that are produced with this
type of stimulation, are almost identical with those of
electro--stimulation, yet many doctors still assume, that they are
two different therapies. This has resulted in TENS being
considered as a daily therapy, while electro--acupuncture or
electro--stimulation were treated as `alternative therapy'. Apart
from the fact, that electro--acupuncture electro--impulses are
delivered through needles, both therapies should be considered
identical. Patients, who have reservations about the use of
needles, can by the use of electric impulses over surface
electrodes on the skin, have a satisfactory alternative (see
Figure \ref{FES}). We choose the term electro--stimulation,
because the acupuncture system is not included in all therapies.
Clinical tests have showed that there are two specific types of
reactions:
\begin{itemize}
\item  The first reaction is spontaneous and dependent on the
choice of body region. The stimulation of this part of the body
results in an unloading, that can be compared with that of a
battery. Normally this goes hand in hand with an immediate
improvement in the patient. This effect of unloading may also be
reached by non--specific electric stimulation.

\item  The second normal reaction is of a delayed nature, that
results in relaxation and control of pain. Moreover two other
important effects follow, that begin between 10 and 20 minutes
after the start of the treatment. This reaction is associated
(combined) with different chemicals, such as beta--endorphins and
5--hydrocytryptamins. When using low and very low frequency
stimulations, the second effect is obtained by the utilization of
specific frequencies on the body. This is independent of the
choice of a specific part of the body, because the connected
electromagnet makes the induction of secondary electric current in
the whole body possible.
\end{itemize}}

In particular, \emph{electro--muscular stimulation} ($EMS$, see
Figure \ref{FES}) represents a union of external electrical
stimulation fields, internal myofibrillar excitation--contraction
paths, and dissipative skin \& fat geometries, formally given by
\begin{equation}
EMS=EMS_{fields}\bigcup EMS_{paths}\bigcup EMS_{geom}. \label{EMS}
\end{equation}
In Feynman's QED formulation, corresponding to each of the three
$EMS$--phases in (\ref{EMS}) we formulate:
\begin{enumerate}
    \item The \textit{least action principle}, to model a unique,
    external--anatomical, predictive and smooth, macroscopic $EMS$ field--path--geometry; and
    \item Associated \textit{Feynman path integral}, to model an ensemble of
    rapidly and stochastically fluctuating, internal, microscopic, fields--paths--geometries
    of the cellular $EMS$, to which the external--anatomical macro--level
    represents both time and ensemble \emph{average}.%
    \footnote{Recall that \textit{ergodic hypothesis}
    equates \textit{time average} with \textit{ensemble average}.}
\end{enumerate}
In the proposed formalism, muscular excitation--contraction paths
$x^i(t)$ are caused by electromagnetic stimulation fields
$\Phi^k=\Phi^k(t,x,y,z)$, by the the Lorenz force equation
(\ref{LorFor}), while they are both affected by dissipative and
noisy skin \& fat shapes and curvatures, defined by the local
Riemannian musculo-skeletal metric tensor $m_{ij}$ (\ref{cov}).
In particular, the electromagnetic 2--form Faraday $F=F_{\alpha
\beta}\,dx_\alpha dx_\beta$, defines the
Lorentz force 1--form $Q=Q_{\alpha}\,dx_\alpha$ of the electro--muscular stimulation,%
\begin{equation*}
Q_{\alpha}\equiv \dot{p}_{\alpha }\propto qF_{\alpha \beta
}v^{\beta},
\end{equation*}%
where $q$ is total electric charge and $v^{\beta}$ is the velocity
vector--field of the stimulation flow. This equation says that the muscular force 1--form $%
Q_{\alpha }$ generated by the simulation is proportional to the
stimulation field strength $F_{\alpha \beta }$, velocity of the
stimulation flow $v^{\beta}$ through the skin--fat--muscle tissue,
as well as the total stimulation charge $q$.

In the following text, we first formulate the global model for the
$EMS$, to set up the general formalism to be specialized
subsequently for each of the three $EMS$--phases.
\begin{figure}[h]
\centerline{\includegraphics[width=12cm]{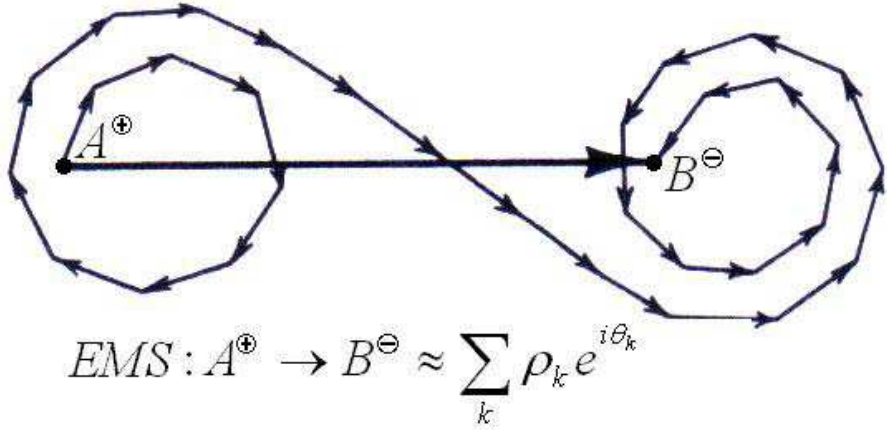}}
\caption{Simplified Feynman--like experimental approach to
electrical muscular stimulation: the flow of electric current from
the positive surface pad $A^\oplus$ to the negative pad
$B^\ominus$ can be approximated by the vector sum of complex
vectors $\rho_k\,{\mathrm e}^{i\theta_k}$, where $\theta_k$ are
proportional to the time taken by each vector $\rho_k$; this
vector sum will be further developed into Feynman
integral.}\label{FeynmanEMS}
\end{figure}

\subsection{Global macro $EMS-$level}

In general, at the \emph{macroscopic} $EMS$--level we first
formulate the \textit{total action} $S[\Phi]$, our central
quantity, which can be described through physical dimensions of
$Energy\times Time=Effort$, which is also the dimension of the
\textit{Planck constant} $\hbar$ ($=1$ in normal units). This
total action quantity has immediate biophysical ramifications:
\emph{the greater the action -- the higher the stimulation effect
on the new shape.} The action $S[\Phi]$ depends on macroscopic
fields, paths and geometries, commonly denoted by an abstract
field symbol $\Phi^i$. The action $S[\Phi]$ is formally defined as
a temporal integral from the \emph{initial} time instant $t_{ini}$
to the \emph{final} time instant $t_{fin}$,
\begin{equation}
S[\Phi]=\int_{t_{ini}}^{t_{fin}}\mathfrak{L}[\Phi]\,dt,
\label{act}
\end{equation}%
with \textit{Lagrangian density}, given by
\begin{equation*}
\mathfrak{L}[\Phi]=\int
d^{n}x\,\mathcal{L}(\Phi^i,\partial_{x^j}\Phi^i),
\end{equation*}%
where the integral is taken over all $n$ coordinates $x^j=x^j(t)$
of the $EMS$, and $\partial_{x^j}\Phi^i$ are time and space
partial derivatives of the $\Phi^i-$variables over coordinates.

Second, we formulate the \textit{least action principle} as a
minimal variation $\delta$ of the action $S[\Phi]$
\begin{equation}
\delta S[\Phi]=0, \label{actPr}
\end{equation}
which, using variational \textit{Euler--Lagrangian equations},
derives field--motion--geometry of the unique and smooth
$EMS-$transition map (or, more appropriately, `functor'
\cite{GaneshADG})
\begin{equation*}
\mathcal{T}:STIMUL_{t_{ini}}\Rightarrow
CONTRACT_{t_{mid}}\Rightarrow SHAPE_{t_{fin}},
\end{equation*}
acting at a macro--level from some initial time $t_{ini}$ to the
final time $t_{fin}$ (via the intermediate time
$t_{mid}$).\footnote{Here, we have in place $n-$\emph{categorical}
\textit{Lagrangian--field structure} on the \textit{muscular
Riemannian configuration manifold} $M$ \cite{GaneshADG}
$$\Phi^{i}:[0,1]\rightarrow M,\,\Phi^{i}:\Phi^{i}_0\mapsto
\Phi^{i}_1,$$ using
$$\frac{d}{dt}f_{\dot{x}^{i}}=f_{x^{i}}
\Rightarrow\partial _{\mu }\left( \frac{\partial
\mathcal{L}}{\partial _{\mu }\Phi^{i}}\right)=\frac{\partial
\mathcal{L}}{\partial \Phi^{i}},$$ with
$$[x_0,x_1]\rightarrowtail[\Phi^{i}_0,\Phi^{i}_1].
$$}

In this way, we get macro--objects in the global $EMS$: a single
electrodynamic stimulation field described by Maxwell field
equations, a single muscular excitation--contraction path
described by Lagrangian equation of motion, and a single
Riemannian skin \& fat geometry.

\subsection{Local Micro $EMS-$level}

After having properly defined macro--level $EMS$, we move down to
the \emph{microscopic cellular} $EMS$--level of rapidly
fluctuating electrodynamic fields, sarcomere--contraction paths
and coarse--grained, fractal muscle--fat geometry, where we cannot
define a unique and smooth field--path--geometry. The most we can
do at this level of \textit{fluctuating noisy uncertainty}, is to
formulate an adaptive path integral and calculate overall
probability amplitudes for ensembles of local transitions from
negative $EMS$--pad $A^\ominus$ to the positive pad $B^\oplus$
(see Figure \ref{FeynmanEMS}). This \textit{probabilistic
transition micro--dynamics} is given by a multi
field--path--geometry, defining the microscopic \textit{transition
amplitude} corresponding to the macroscopic $EMS-$transition map
$\mathcal{T}$. So, what is externally the transition map,
internally is the transition amplitude. The absolute square of the
transition amplitude is the \textit{transition probability}.

Now, the total $EMS-$transition amplitude, from the initial state
$STIMUL$, to the final state $SHAPE$, is defined on $EMS$
\begin{eqnarray}\langle SHAPE|STIMUL\rangle:
STIMUL_{t_{0}}\Rrightarrow SHAPE_{t_{1}}, \label{transdyn}
\end{eqnarray} given by \emph{modern adaptive
generalization} of the Feynman's path integral, see
\cite{QuLeap}). The transition map (\ref{transdyn}) calculates
\textit{overall probability amplitude} along a multitude of wildly
fluctuating fields, paths and geometries, performing the
\emph{microscopic} transition from the micro--state
$STIMUL_{t_{0}}$ occurring at initial micro--time instant $t_{0}$
to the micro--state $SHAPE_{t_{1}}$ at some later micro--time
instant $t_{1}$, such that all micro--time instants fit inside the
global transition interval $t_0,t_1,...,t_s\in[t_{ini},t_{fin}]$.
It is symbolically written as
\begin{equation}
\langle SHAPE|STIMUL\rangle =\int\mathcal{D}[w\Phi]\, {\mathrm
e}^{\mathrm i S[\Phi]}, \label{pathInt}
\end{equation}
where the Lebesgue integration is performed over all continuous
$\Phi^i_{con}=fields+paths+geometries$, while summation is
performed over all discrete processes and regional topologies
$\Phi^j_{dis}$. The symbolic differential $\mathcal{D}[w\Phi]$ in
the general path integral (\ref{pathInt}), represents an
\textit{adaptive path measure}, defined as a weighted product
\begin{equation}
\mathcal{D}[w\Phi]=\lim_{N\to\infty}\prod_{s=1}^{N}w_sd\Phi _{s}^i, \qquad ({%
i=1,...,n=con+dis}),  \label{prod}
\end{equation}
which is in practice satisfied with a large $N$.

In this way, we get a range of micro--objects in the local $EMS$
at the short time--level: ensembles of rapidly fluctuating, noisy
and crossing electrical stimulation fields, myofibrillar
contraction paths and local skin \& fat shape--geometries.
However, by averaging process,
 both in time and along ensembles of fields, paths and geometries, we can
 recover the corresponding global, smooth and fully predictive,
 external $EMS$ transition--dynamics $\mathcal{T}$.

\subsection{Micro--Level Adaptation and Muscular Training}

The adaptive path integral (\ref{pathInt}--\ref{prod})
incorporates the local muscular training process according to the
basic learning formula
\[
NEW\,\, VALUE = OLD\,\, VALUE + INNOVATION,
\]
where the term $VALUE$ represents respectively \textit{biological
images} of the $STIMUL$,\\ $CONTRACT$ and $SHAPE$.

The general \textit{synaptic weights} $w_s=w_s(t)$ in (\ref{prod})
are updated by the \textit{homeostatic neuro--muscular feedbacks}
during the transition process $\mathcal{T}$, according to one of
the two standard neural training schemes, in which the micro--time
level is
traversed in discrete steps, i.e., if $t=t_0,t_1,...,t_s$ then $%
t+1=t_1,t_2,...,t_{s+1}$:

\begin{enumerate}
\item  A \textit{self--organized}, \textit{unsupervised}, e.g.,
Hebbian--like training rule (see, e.g. \cite{NeuFuz}:
\begin{equation}
w_s(t+1)=w_s(t)+ \frac{\sigma}{\eta}(w_s^{d}(t)-w_s^{a}(t)),
\label{Hebb}
\end{equation}
where $\sigma=\sigma(t),\,\eta=\eta(t)$ denote \textit{signal} and \textit{%
noise}, respectively, while superscripts $d$ and $a$ denote
\textit{desired} and \textit{achieved} muscular micro--states,
respectively; or

\item  A certain form of a \emph{supervised gradient descent
training}:
\begin{equation}
w_s(t+1)\,=\,w_s(t)-\eta \nabla J(t),  \label{gradient}
\end{equation}
where $\eta $ is a small constant, called the \textit{step size},
or the \emph{training rate,} and $\nabla J(n)$ denotes the
gradient of the `performance hyper--surface' at the $t-$th
iteration.
\end{enumerate}
For more details EMS--fields, paths and geometries, see
\cite{QuLeap}.

\section{Appendix: Manifolds and Bundles}

\subsection{Manifolds}

\begin{figure}[h]
\centerline{\includegraphics[width=8cm]{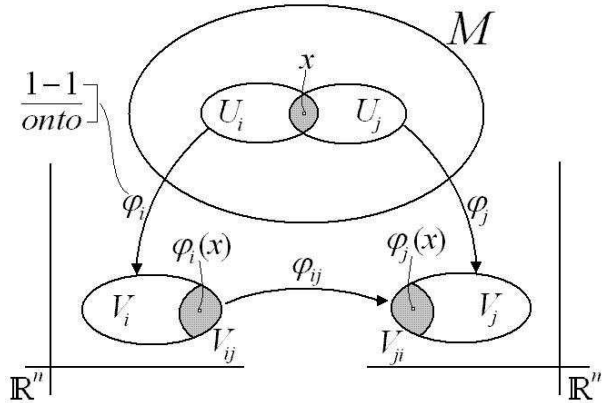}}
\caption{Geometric picture of the manifold concept.}
\label{Manifold1}
\end{figure}

Smooth manifold is a curved $n$D space which is locally equivalent
to $\mathbb{R}^n$. To sketch it formal definition, consider a set
$M$ (see Figure \ref{Manifold1}) which is a \emph{candidate} for a
manifold. Any point $x\in M$\footnote{Note that sometimes we will
denote the point in a manifold $M$ by $m$, and sometimes by $x$
(thus implicitly assuming the existence of coordinates
$x=(x^i)$).} has its \textit{Euclidean chart}, given by a 1--1 and
\emph{onto} map $\varphi _{i}:M\rightarrow \Bbb{R}^{n}$, with its
\textit{Euclidean image} $V_{i}=\varphi_{i}(U_{i})$. More
precisely, a chart $ \varphi_{i}$ is defined by
\[
\varphi _{i}:M\supset U_{i}\ni x\mapsto \varphi _{i}(x)\in V_{i}\subset \Bbb{%
R}^{n},
\]
where $U_{i}\subset M$ and $V_{i}\subset \Bbb{R}^{n}$ are open
sets.

Clearly, any point $x\in M$ can have several different charts (see
Figure \ref{Manifold1}). Consider a case of two charts, $\varphi
_{i},\varphi _{j}:M\rightarrow \Bbb{R}^{n}$,
having in their images two open sets, $V_{ij}=\varphi _{i}(U_{i}\cap U_{j})$ and $%
V_{ji}=\varphi _{j}(U_{i}\cap U_{j})$. Then we have
\textit{transition functions} $\varphi _{ij}$ between them,
\[
\varphi _{ij}=\varphi _{j}\circ \varphi
_{i}^{-1}:V_{ij}\rightarrow V_{ji},\qquad \text{locally given
by\qquad }\varphi _{ij}(x)=\varphi _{j}(\varphi _{i}^{-1}(x)).
\]
If transition functions $\varphi _{ij}$ exist, then we say that
two charts, $\varphi _{i}$ and $\varphi _{j}$ are
\emph{compatible}. Transition functions represent a general
(nonlinear) \emph{transformations of coordinates}, which are the
core of classical \emph{tensor calculus}.

A set of compatible charts $\varphi _{i}:M\rightarrow
\Bbb{R}^{n},$ such that each point $x\in M$ has its Euclidean
image in at least one chart, is called an \textit{atlas}. Two
atlases are \emph{equivalent} iff all their charts are compatible
(i.e., transition functions exist between them), so their union is
also an atlas. A \textit{manifold structure} is a class of
equivalent atlases.

Finally, as charts $\varphi _{i}:M\rightarrow \Bbb{R}^{n}$ were
supposed to be 1-1 and onto maps, they can be either
\emph{homeomorphism}\emph{s}, in which case we have a
\emph{topological} ($C^0$) manifold, or
\emph{diffeomorphism}\emph{s}, in which case we have a
\emph{smooth} ($C^{k}$) manifold.

\subsection{Fibre Bundles}

\begin{figure}[h]
 \centerline{\includegraphics[width=12cm]{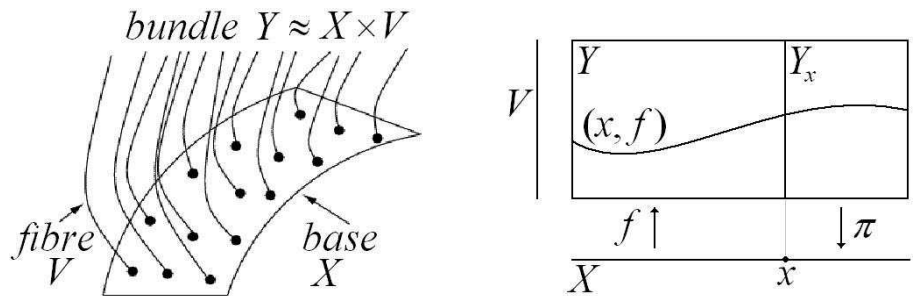}}
\caption{A sketch of a fibre bundle $Y\approx X\times V$ as a
generalization of a product space $X\times V$; left -- main
components; right -- a few details (see text for
explanation).}\label{Fibre1}
\end{figure}

On the other hand, the well--known tangent bundle $TM$ and
cotangent bundle $T^{\ast }M$ of a smooth manifold $M$,
respectively endowed with Riemannian geometry (suitable for
Lagrangian dynamics) and symplectic geometry (suitable for
Hamiltonian dynamics) -- are special cases of a more general
geometrical object called \emph{fibre bundle}. Here the word
\emph{fiber} $V$ of a map $\pi :Y\rightarrow X$ denotes the
\emph{preimage} $\pi^{-1}(x)$ of an element $x\in X$. It is a
space which \emph{locally} looks like a product of two spaces
(similarly as a manifold locally looks like Euclidean space), but
may possess a different \emph{global} structure. To get a visual
intuition behind this fundamental geometrical concept, we can say
that a fibre bundle $Y$ is a \emph{homeomorphic generalization} of
a \emph{product space} $X\times V$ (see Figure \ref{Fibre1}),
where $X$ and $V$ are called the \emph{base} and the \emph{fibre},
respectively. $\pi :Y\rightarrow X$ is called the
\emph{projection}, $Y_{x}=\pi ^{-1}(x)$ denotes a fibre over a
point $x$ of the base $X$, while the map $f=\pi ^{-1}:X\rightarrow
Y$ defines the \emph{cross--section}, producing the \textit{graph}
$(x,f(x))$ in the bundle $Y$ (e.g., in case of a tangent bundle,
$f=\dot{x}$ represents a velocity vector--field, so that the graph
in a the bundle $Y$ reads $(x,\dot{x})$).

The main reason why we need to study fibre bundles is that
\emph{all dynamical objects} (including vectors, tensors,
differential forms and gauge potentials) are their
\emph{cross--sections}, representing \emph{generalizations of
graphs of continuous functions}.

\end{document}